\documentclass{aa}
\usepackage{xcolor}
\usepackage{natbib,twoopt}
\usepackage{graphicx}
\usepackage{txfonts}
\usepackage{lastpage}
\usepackage[breaklinks]{hyperref}
\usepackage{orcidlink}

\bibpunct{(}{)}{;}{a}{}{,}
\definecolor{cobalt}{rgb}{0.06, 0.2, 0.65}
\hypersetup{
  colorlinks,
  citecolor=cobalt,
  linkcolor=[rgb]{0.8, 0.2, 1.0},
  urlcolor=cobalt,
}

\begin{document} 

\newcommand{\Msun}{\mathrm{M}_{\odot}}
\newcommand{\Mh}{M_{\mathrm{h}}}
\newcommand{\LCDM}{\Lambda\text{CDM}}
\newcommand{\Mstar}{M_{\star}}
\newcommand{\hMpc}{\ensuremath{h^{-1}\,\mathrm{Mpc}}}
\newcommand{\hMpcden}{\ensuremath{h^{3}\,\mathrm{Mpc}^{-3}}}
\newcommand{\hMsun}{\ensuremath{h^{-1}\,\mathrm{M}_{\odot}}}
\newcommand{\Abs}[1]{\left\lvert\mathbf{#1}\right\rvert}

\title{On the dependence of galaxy assembly bias on the selection criteria, number density, and redshift of galaxy samples}

\titlerunning{Galaxy assembly bias across galaxy samples}

\author{S.~García-Moreno
      \inst{1}\thanks{\email{segarc19@ucm.es}}\orcidlink{0009-0005-5691-894X}
      \and
      J.~Chaves-Montero\inst{2}\thanks{\email{jchaves@ifae.es}}\orcidlink{0000-0002-9553-4261}    
      }

\authorrunning{S.~García-Moreno \& J.~Chaves-Montero}

\institute{Departamento de Física de la Tierra y Astrofísica, Fac. de C.C. Físicas, Universidad Complutense de Madrid, E-28040 Madrid, Spain
     \and
         Institut de Física d’Altes Energies (IFAE), The Barcelona Institute of Science and Technology, 08193 Bellaterra (Barcelona), Spain}

\date{Received 9 April 2025; accepted 5 January 2026}

\abstract
{
One of the key factors influencing galaxy clustering in the nonlinear regime is galaxy assembly bias, which describes the dependence of galaxy clustering on halo properties beyond halo mass. We study this effect by analyzing galaxy samples selected according to stellar mass, luminosity, and broad-band colors from the IllustrisTNG hydrodynamical simulation. We find that galaxy assembly bias depends strongly upon the selection criteria, number density, and redshift of the galaxy sample, with this effect increasing or decreasing galaxy clustering by as much as 25\%. Interestingly, no single secondary halo property fully captures the strength of galaxy assembly bias for any galaxy population. Therefore, empirical models predicting galaxy assembly bias as a function of a single halo property cannot reproduce predictions from hydrodynamical simulations. Finally, we investigate how galaxy assembly bias arises from the interplay between halo assembly bias --- the dependence of halo clustering on properties other than halo mass --- and occupancy variation --- the correlation between galaxy occupation and secondary halo properties. We provide a fast analytical expression to predict the level of galaxy assembly bias induced by any halo property in simulated galaxy catalogs without the need for computationally expensive shuffling techniques.}

\keywords{large-scale structure of Universe --
                Galaxies: formation --
                Galaxies: statistics --
                Cosmology: theory
               }

\maketitle
%
%________________________________________________________________

\section{Introduction}

Structure formation theories predict that galaxies form as gas cools and condenses within virialized dark matter structures known as halos \citep{TwoStageGalaxyFormationWhiteRees1978}. Consequently, the formation, growth, and properties of galaxies are likely to be closely connected to the growth, internal properties, and distribution of halos. Small-scale measurements of galaxy clustering and galaxy-galaxy lensing are highly sensitive to this relationship, the galaxy-halo connection, making its thorough understanding crucial for cosmological analyses \citep{chavesGalFormationOriginLensingLow2023, contrerasbosslil2023}.

Of the multiple effects affecting the galaxy-halo connection, we examine the dependence of galaxy clustering on halo properties beyond mass, an effect commonly known as galaxy assembly bias \citep{crotonHaloAssemblyBias2007a, zentnerGalaxyAssemblyBias2014a}. This effect is a prevalent prediction of the most advanced hydrodynamical simulations, including EAGLE \citep{chaves-monteroSubhaloAbundanceMatching2016}, Illustris \citep{artaleImpactAssemblyBias2018a}, IllustrisTNG \citep{hadzhiyska2021GalaxyHaloConnectionIllustrisTNG}, and MillenniumTNG \citep{contreras2023MilleniumTNG}, as well as semi-analytic models such as the precursor of SAGE \citep{crotonHaloAssemblyBias2007a, contreras2021FlexibleModellingGalaxy}, L-GALAXIES \citep{zehaviImpactAssemblyBias2018a, contrerasEvolutionAssemblyBias2019a}, Semi-Analytical Galaxies \citep{jimenez2020DistributionDarkGalaxies}, and Santa Cruz \citep{hadzhiyska2021GalaxyBiasIllustrisTNGSantaCruz}. However, the strength of this effect depends upon the properties of the galaxy sample and, for a particular galaxy sample, varies across galaxy formation models.

Detecting galaxy assembly bias in observations is inherently challenging, as a direct detection would require unambiguously associating each galaxy with its host dark matter halo, and accurately estimating the mass of such halos. Both steps are difficult: assigning galaxies to halos is uncertain, and measuring individual halo masses at cosmological distances remains highly nontrivial \citep[e.g.,][]{zhao2025_halomass}. Consequently, most observational analyses infer galaxy assembly bias indirectly by fitting empirical models to galaxy clustering data, where specific parameters govern the strength of assembly bias. A robust detection therefore requires an empirical model sufficiently flexible to capture all relevant astrophysical effects influencing galaxy clustering from intermediate to small scales; otherwise, the impact of these effects may be mistakenly interpreted as evidence of galaxy assembly bias. There is considerable disagreement in the literature regarding the detection of galaxy assembly bias, with some studies reporting positive detections \citep[e.g.,][]{Yuan2021_gab, Wang2022_gab, Yuan2022_yesgab, contrerasbosslil2023, Alam2024_gab, Paviot2024_gab, Pearl2024_gab}, while others find no conclusive evidence for its presence \citep[e.g.,][]{zu2018MappingStellarContent, niemiec2018GalaxyBiasLRGweakLensing, yuan2020_lilab, McCarthy2022_gab, Yuan2022_nogab}. These inconsistencies are partly due to the fact that many analyses constrain only the model parameters governing assembly bias, rather than quantifying its percent-level effect on clustering, which complicates interpretation because the link between parameter values and the actual magnitude of assembly bias is often nontrivial.

The primary goal of this work is to systematically quantify galaxy assembly bias for the galaxy populations most representative of large-scale structure surveys. Spectroscopic surveys such as the Sloan Digital Sky Survey \citep[SDSS;][]{york2000_sdss} and the Dark Energy Spectroscopic Instrument \citep[DESI;][]{DESI2016a.Science} typically target luminous red galaxies (LRGs) at intermediate redshift, emission line galaxies (ELGs) at higher redshift, and stellar-mass–selected samples at low redshift. In contrast, weak-lensing surveys such as the Dark Energy Survey \citep[DES;][]{des_red2005} and the Kilo-Degree Survey \citep[KiDS;][]{kids2021} generally select galaxies based on their brightness in a broad band. In this work, we quantify galaxy assembly bias as a function of redshift and number density for samples selected by stellar mass, r-band magnitude, and broad-band colors, designed to emulate the aforementioned observational selection criteria.

To this end, we analyze the largest hydrodynamical simulation of the IllustrisTNG suite \citep{springelFirstResultsIllustrisTNG2018a, marinacciFirstResultsIllustrisTNG2018a, naimanFirstResultsIllustrisTNG2018a, pillepichFirstResultsIllustrisTNG2018a, nelsonFirstResultsIllustrisTNG2018a}, which accurately reproduces the clustering of red and blue galaxies as a function of redshift, in good agreement with SDSS/DR7 observations across a wide range of stellar masses \citep{springelFirstResultsIllustrisTNG2018a}. We therefore expect its predictions for galaxy assembly bias to be more representative of the real Universe than those from other hydrodynamical or semi-analytic models that exhibit significant discrepancies in galaxy clustering. Consequently, this work presents the first systematic study of galaxy assembly bias across the main galaxy populations relevant for large-scale structure analyses, all within a galaxy formation model describing the clustering of these populations accurately.

In addition to measuring the level of galaxy assembly bias for these samples, we study how galaxy assembly bias emerges from the interplay of halo assembly bias \citep[e.g.,][]{gaoAgeDependenceHalo2005a, wechslerDependenceHaloClustering2006, gaoAssemblyBiasClustering2007} --- the dependence of halo clustering on properties other than halo mass --- and occupancy variation \citep{zehaviImpactAssemblyBias2018a} --- the correlation between galaxy occupation and secondary halo properties. We present a fast and practical technique for estimating the level of galaxy assembly bias for a galaxy population associated with a given secondary halo property, relying solely on the median value of this property for the host halos of these galaxies and the level of halo assembly bias as a function of the target property. We validate this method using data from the IllustrisTNG simulation. The precision of this approach can be further improved by measuring halo assembly bias from a much larger gravity-only simulation, as halo assembly bias is largely independent of cosmology \citep{contrerasEvolutionAssemblyBias2019a}.

This paper is organized as follows. In Section \ref{sec:data}, we present the IllustrisTNG simulation and extract multiple galaxy samples from it. We measure the strength of galaxy assembly bias from these samples in Section \ref{sec_GAB_measure}, and we study the emergence of this effect from the interplay of halo assembly bias and occupancy variation in Section \ref{sec_GAB_predict}. We summarize our main findings and conclude in Section \ref{sec_Conclusions}.

\section{Simulation} \label{sec:data}

In Sections \ref{sec:data_tng300}, \ref{sec:data_samples}, and \ref{sec:data_hprop}, we introduce the IllustrisTNG simulation, describe the galaxy samples extracted from this simulation, and outline the halo properties used to study galaxy assembly bias, respectively.

\subsection{IllustrisTNG simulation}
\label{sec:data_tng300}

In this work we analyze galaxy samples extracted from the TNG300-1 hydrodynamical simulation, the largest from the IllustrisTNG suite\footnote{\url{https://www.tng-project.org/}} \citep{springelFirstResultsIllustrisTNG2018a, marinacciFirstResultsIllustrisTNG2018a, naimanFirstResultsIllustrisTNG2018a, pillepichFirstResultsIllustrisTNG2018a, nelsonFirstResultsIllustrisTNG2018a}. This simulation was performed using the moving-mesh code \texttt{AREPO} \citep{springelPurSiMuove2010a}, which tracks the joint evolution of dark matter, gas, stars, and supermassive black holes employing a comprehensive galaxy formation model \citep{weinbergerSimulatingGalaxyFormation2017a, pillepichSimulatingGalaxyFormation2018a}.

The TNG300-1 simulation evolved $2500^3$ gas cells and an equal number of dark matter particles within a periodic box of $205\,\hMpc$ on a side adopting the Planck 2015 cosmology \citep{planckcollaborationPlanck2015Results2016a}. The initial masses of the gas tracers and dark matter particles were $0.7 \times 10^7$ and $4.0 \times 10^7\,\hMsun$, respectively. Halos were detected using a standard friends-of-friends group finder with a linking length of $b=0.2$ \citep{davisEvolutionLargescaleStructure1985}, while self-bound substructures within halos, commonly known as subhalos, were identified with the \texttt{SUBFIND} algorithm \citep{springelPopulatingClusterGalaxies2001b, dolagSubstructuresHydrodynamicalCluster2009}.

It is standard to refer to subhalos located at the potential minimum of their host halos as centrals, while other subhalos are designated as satellites. Central and satellite subhalos hosting a stellar component are identified as central or satellite galaxies, respectively. Some subhalos in the catalog were erroneously identified as independent halos by \texttt{SUBFIND}, leading to the misclassification of their galaxies as centrals instead of satellites. These subhalos are typically known as backsplash subhalos. We leverage merger tree information to mitigate this problem, reclassifying backsplash subhalos as satellites and assigning these to the correct host halo.

We use the \texttt{LHaloTree} merger trees \citep{SpringelSimulationsGalaxiesandQuasars2005} to reclassify central subhalos at $z_i$ as satellites if these were identified as satellites for at least five snapshots from $z=3$ to $z_i$, where $z_i=0$, 0.5, 1, and 2 are the four snapshots we analyze in this work. The five-snapshot threshold was chosen to mitigate transient tracking artifacts inherent to merger tree algorithms \citep[e.g.,][]{kong_bloodhound_2025}. Finally, we associate the reclassified subhalos with the last halos these interacted with, provided that the target halo mass exceeds that of the subhalo.

\subsection{Galaxy samples} \label{sec:data_samples}

\begin{figure}
    \centering
    \includegraphics[width=\linewidth]{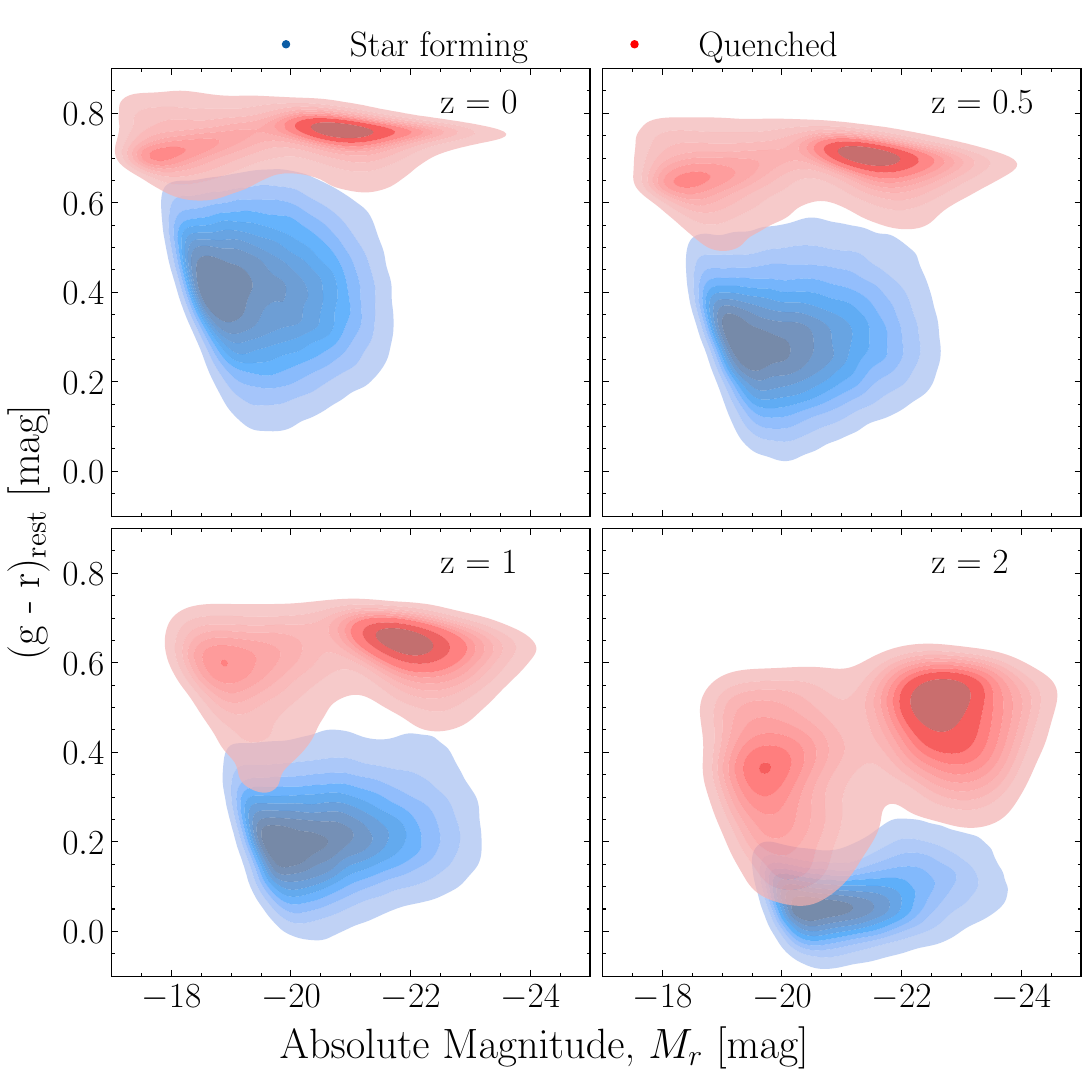}
    \caption{
    Rest-frame color-magnitude diagram of TNG-300 galaxies with $\Mstar>10^{9}h^{-1}\Msun$. Blue and red colors display the results for galaxies with specific star formation rates greater and smaller than $\log_{10}({\rm sSFR} [{\rm yr}^{-1}]) = -11$, respectively. Contours denote deciles of the populations, with darker shaded areas indicating the most densely populated regions.
    }
    \label{fig_color_magnitude_quenched}
\end{figure}

\begin{figure*}
    \centering
    \includegraphics[width=0.475\linewidth]{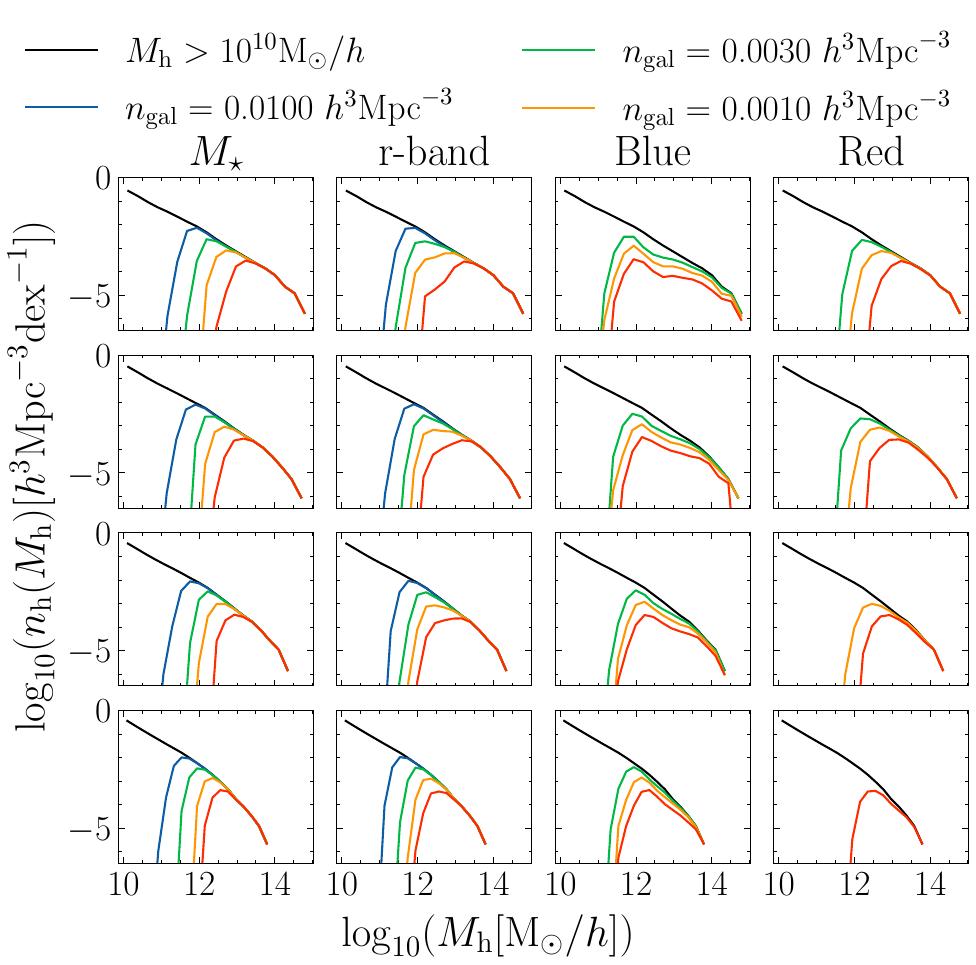}\includegraphics[width=0.475\linewidth]{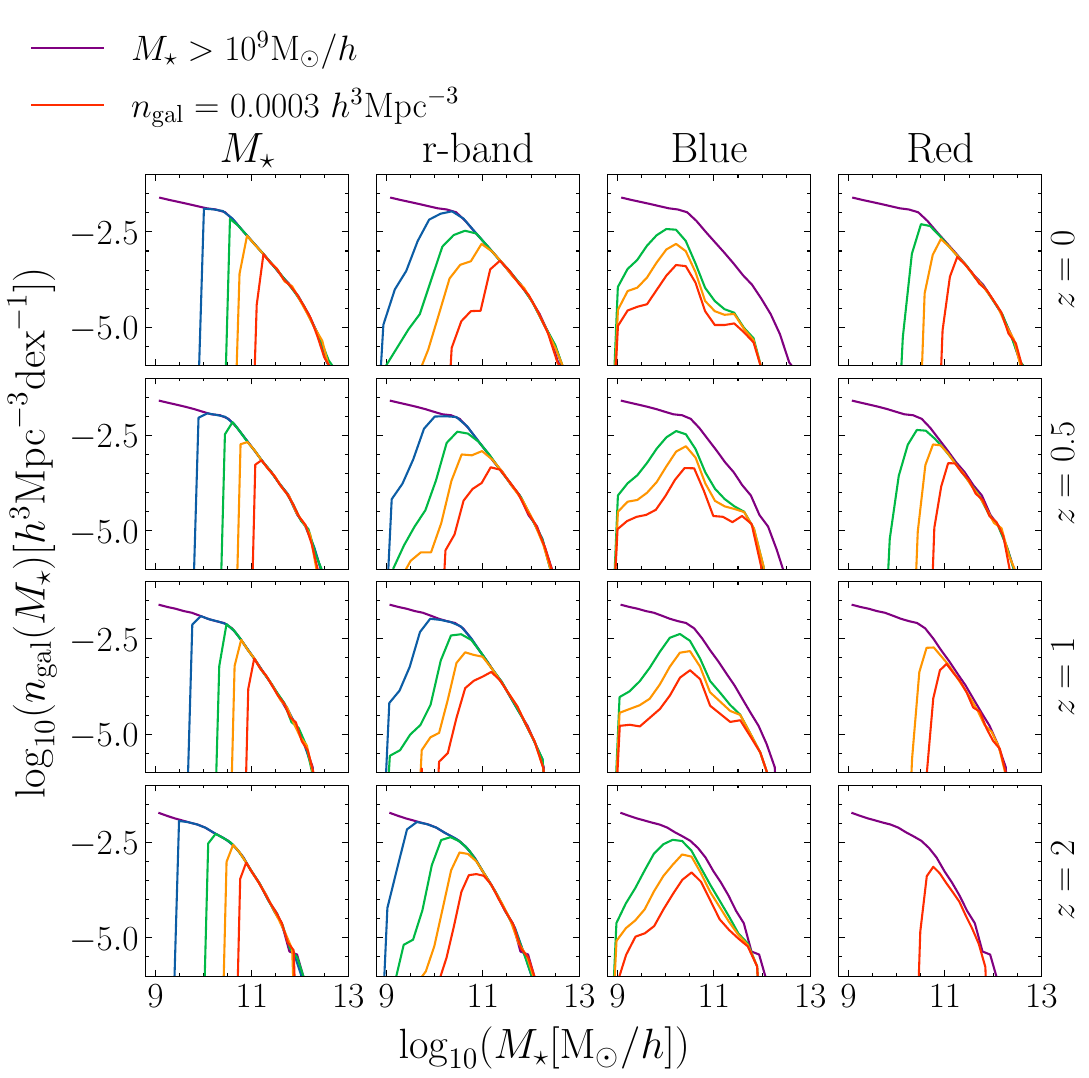}
    \caption{
    Halo mass function (left panels) and stellar mass function (right panels) for the stellar mass, $r$-band, blue, and red samples. Each column corresponds to a different sample, while rows represent results at different redshifts. Colored lines show the mass functions for samples with distinct number densities. The black lines in the left panels display the halo mass function for $\Mh > 10^{10} \, h^{-1} \, \mathrm{M_\odot}$ halos, while the purple lines in the right panels depict the stellar mass function $M_\star > 10^{9} \, h^{-1} \, \mathrm{M_\odot}$ galaxies.}
    \label{fig_halo_stelar_mass_function}
\end{figure*}

We analyze galaxy assembly bias using galaxy samples selected according to stellar mass, $r$-band magnitude, and two samples based on broad-band colors: blue and red. Specifically, we use stellar masses measured from all star particles bound to a subhalo and magnitudes resulting from the added luminosity of all these particles. We create versions of these four samples with number densities $n = 0.01$, $0.003$, $0.001$, and $0.0003\, \hMpcden$ at $z = 0$, $0.5$, $1$, and $2$. To produce each of the stellar mass and r-band samples, we select galaxies with the greatest value of the corresponding property until reaching the target number density. We generate the blue and red samples by first selecting all galaxies in the position of the color-magnitude diagram roughly corresponding to galaxies with specific star formation rates greater and smaller than $\log_{10}({\rm sSFR} [{\rm yr}^{-1}]) = -11$, respectively, see Fig.~\ref{fig_color_magnitude_quenched}. Then, for the blue and red samples, we select the galaxies with the greatest star formation rate and $r$-band magnitude until reaching the target number density, respectively.

We only use galaxies with stellar mass greater than $M_{\star} = 10^{9} \, \hMsun$, equivalent to more than 100 star particles, to generate these samples in order to ensure that all galaxies are sufficiently resolved. Furthermore, we do not produce the blue sample with $n = 0.01\, \hMpcden$ due to the limited resolution of the simulation, nor the red samples with $n = 0.01\, \hMpcden$ at all redshifts, $n = 0.003\, \hMpcden$ at $z = 1$ and $2$, and $n = 0.001\, \hMpcden$ at $z = 2$ owing to its limited volume. In the left and right panels of Figure \ref{fig_halo_stelar_mass_function}, we show the halo and stellar mass functions of the 53 resulting samples, respectively. The host halos of all galaxies in the samples have masses greater than $\Mh = 10^{10} \, \hMsun$, corresponding to approximately 250 particles, which ensures that these are well resolved.

\subsection{Halo samples} \label{sec:data_hprop}

We measure halo assembly bias for concentration, spin, and formation time using halos with masses greater than $\Mh = 10^{10} \, h^{-1} \, \mathrm{M_\odot}$ at $z = 0$, $0.5$, $1$, and $2$. In what follows, we describe these properties.

We estimate the concentration of dark matter halos as the ratio of the maximum circular velocity to $V_{200} = G^{1/2} M_{200}^{1/2} R_{200}^{-1/2}$ \citep{gaoAssemblyBiasClustering2007}, where $M_{200}$ is the halo mass enclosed within a sphere of mean density 200 times the critical density of the Universe, $R_{200}$ is the radius of this sphere, and $G$ is the gravitational constant. Although more precise methods for estimating halo concentration exist \citep[e.g.,][]{childHaloProfilesConcentration2018}, our approach provides sufficient accuracy for ranking halos by concentration.

Following \citet{bullockUniversalAngularMomentum2001a}, we compute the halo spin as 
\begin{equation}
    \lambda = \frac{J_{200}}{\sqrt{2} M_{200} V_{200} R_{200}},
\end{equation}
where $J_{200}$ is the magnitude of the angular momentum within a sphere of radius $R_{200}$. While the number of particles required to compute the halo spin accurately exceeds the 250-particle limit we impose, the resulting estimates, though noisy, are sufficiently precise for ranking halos according to this property.

Finally, we define the halo formation time, $t_\mathrm{f}$, as the cosmic lookback time at which a halo first attains half of its mass at each of the 4 redshifts used in this work. This property is measured from the TNG300 merger trees.

\section{Measurements of galaxy assembly bias}
\label{sec_GAB_measure}

\begin{figure}
    \centering
    \includegraphics[width=0.975\columnwidth]{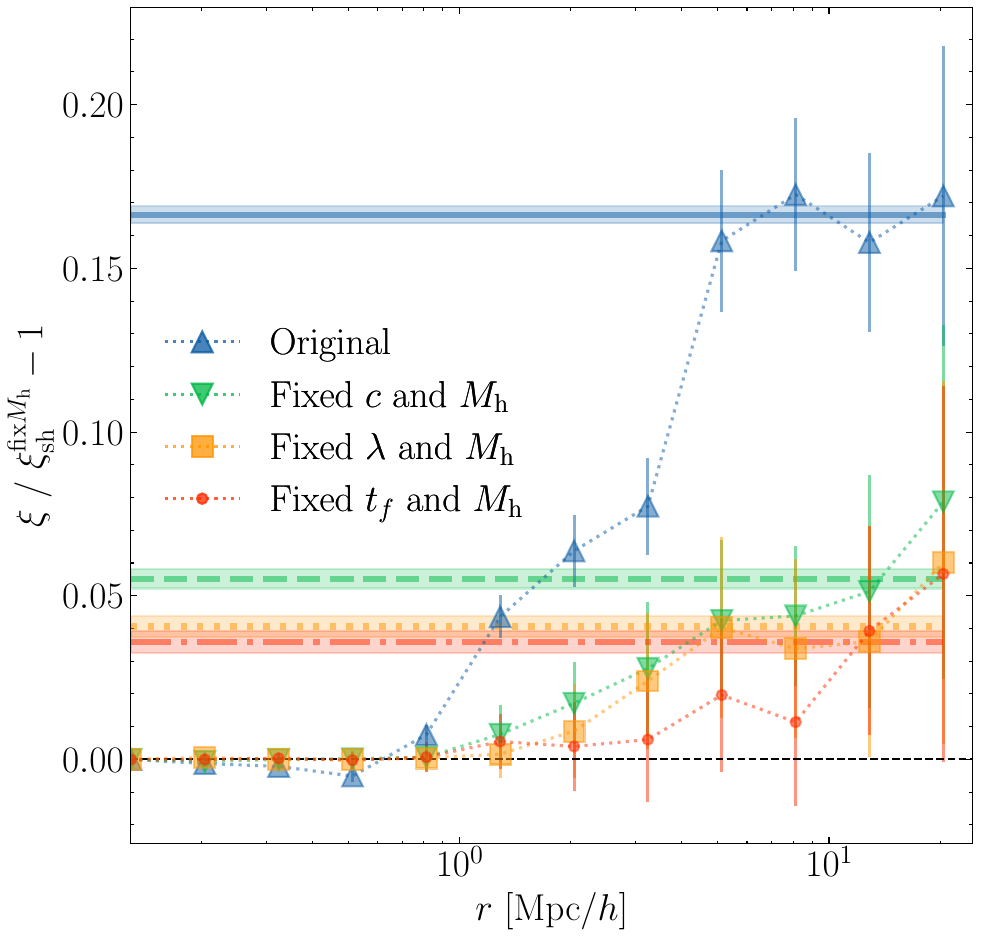}
    \caption{
    Galaxy assembly bias for the stellar mass sample with number density $n = 0.003\ \hMpcden$ at $z=0$. 
    The blue symbols show the ratio of the clustering of this sample and of a modified version where galaxies are shuffled among halos of the same mass. The green, orange and red symbols show this ratio for modified versions of this sample where galaxies are shuffled among halos of the same mass and concentration, spin, and formation time, respectively. Horizontal lines display the average ratio on large scales, which corresponds to the level of galaxy assembly bias. Error bars and shaded areas indicate $1\sigma$ uncertainties.
    }
    \label{fig_GAB_instantaneous}
\end{figure}

\begin{figure*}
    \centering
    \includegraphics[width=0.8\textwidth]{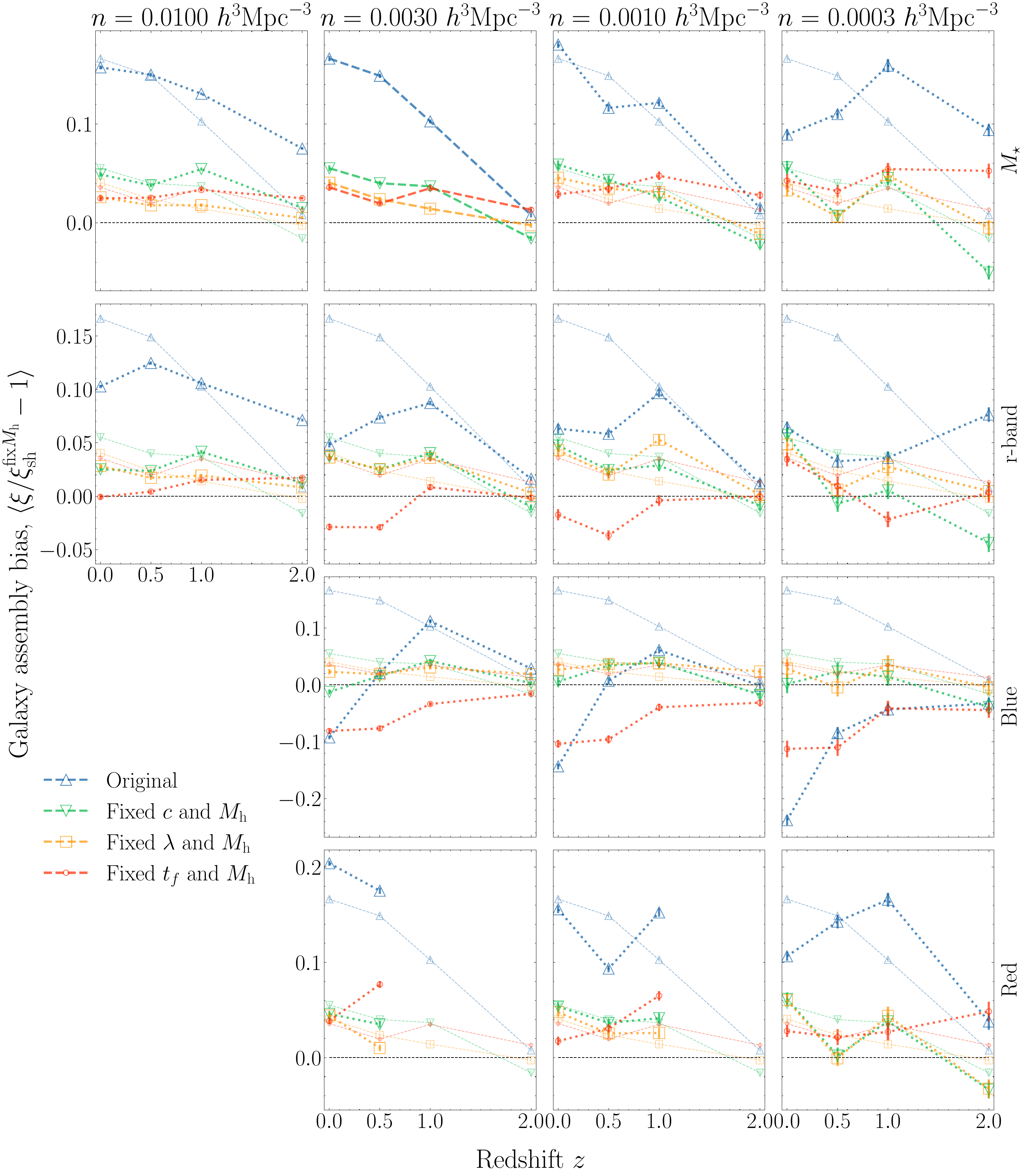}
    \caption{
    Measurements of galaxy assembly bias captured by each secondary property from the stellar mass, $r$-band, blue, and red galaxy samples (rows) across different number densities (columns) as a function of redshift. The y-axis displays our measurements of galaxy assembly bias, which we extract by averaging from $8$ to $25 \hMpc$ the ratio of galaxy clustering of the sample shown in the legend and that of a modified version where galaxies are shuffled among halos of the same mass. In all panels, for ease of comparison, symbols connected by dashed lines represent results for the stellar mass sample with $n = 0.003\, \hMpcden$. The level of galaxy assembly bias varies significantly across samples, redshifts, and number density, increasing or decreasing galaxy clustering by as much as $\simeq24\%$.
    }
    \label{fig_GAB}
\end{figure*}

We measure the level of galaxy assembly bias from simulated galaxy samples using the "shuffling" technique proposed by \citet{crotonHaloAssemblyBias2007a}. This method involves exchanging the galaxies hosted by each dark matter halo among halos of similar masses, effectively removing any clustering dependence on halo properties beyond halo mass. The level of galaxy assembly bias is quantified by comparing the clustering of the original galaxy sample with that of the shuffled sample. In what follows, we give more details about the shuffling procedure and present the level of galaxy assembly bias for each of the samples introduced in Sect. \ref{sec:data_samples}.

We begin by dividing all halos into logarithmic halo mass bins with width $\Delta \log_{10}(\Mh[\hMsun]) = 0.15$. We checked that using bins with slightly different sizes does not alter the results. We then calculate the relative distance between each galaxy and the center of potential of its host. Next, we shuffle the galaxies of all halos within each halo mass bin while keeping the relative distances of galaxies to the halo center fixed, therefore preserving the 1-halo term. Halos without galaxies are also included in the shuffling process. After that, we compute the two-point correlation function of both the unshuffled and shuffled samples, $\xi$ and $\xi_\mathrm{sh}^{\mathrm{fix}\, \Mh}$.

We employ the \texttt{corrfunc} \citep{sinhaCorrfuncSuiteBlazing2020a} package to compute the two-point correlation function, using 12 logarithmically spaced radial bins of width $\Delta \log_{10}(r[h^{-1} {\rm Mpc}])= 0.2$ between $r=0.1$ and $25\ h^{-1} {\rm Mpc}$. To mitigate noise introduced by the stochastic nature of the shuffling technique, we compute 100 shufflings with different random seeds. We then calculate the median ratio between the clustering of the original sample and each shuffled catalog, $\xi/\xi_\mathrm{sh}^{\mathrm{fix}\, \Mh}$, on scales from $r =8$ to $25\ h^{-1} {\rm Mpc}$. Finally, we estimate the level of galaxy assembly bias, along with its uncertainty, by computing the mean and standard deviation of the results for the 100 shufflings. We verified that further increasing the number of shufflings does not improve the results, and excluded larger scales in the calculation to minimize the influence of numerical artifacts due to the simulation's finite volume.

In Fig.~\ref{fig_GAB_instantaneous}, we display the level of galaxy assembly bias for galaxies selected by stellar mass with $n=0.003\ h^3{\rm Mpc}^{-3}$ at $z=0$. Blue symbols and error bars display the mean and standard deviation of the ratio between the clustering of the original and shuffled samples. This small-scale ratio approaches unity since the shuffling procedure preserves the relative distance between satellite galaxies, increases from $r=1$ to $\simeq5\,\hMpc$, and remains nearly constant for larger scales. We find a similar trend for other samples, number densities, and redshifts. This trend justifies using scales larger than $5\,\hMpc$ to measure the level of galaxy assembly bias, which is indicated by the blue horizontal line. The horizontal line is above unity, below, and at the same level if galaxy assembly bias is positive, negative, and null for a particular sample.

We then measure the level of galaxy assembly bias associated with halo concentration, spin, and formation time (see Sect. \ref{sec:data_hprop} for the definitions of these properties). Within each halo mass bin, we further subdivide halos into up to ten bins according to the selected secondary property, reducing the number of sub-bins only when a mass bin contains fewer than ten halos. We then shuffle the galaxy populations among halos within each sub-bin, effectively fixing both halo mass and the targeted secondary property \citep{xuDissectingModellingGalaxy2021a}. We verified that the results are stable against moderate variations in the number of sub-bins.

In Fig.~\ref{fig_GAB_instantaneous}, green, orange, and red symbols show the ratio between the clustering of samples shuffled while holding fixed halo mass and concentration, spin, and formation time, respectively, and samples shuffled while only holding fixed halo mass. We display the results for the average of 100 shufflings. Colored horizontal lines indicate the galaxy assembly bias captured by each property; we compute these following the same approach as for computing the total assembly bias of the sample. If the galaxy assembly bias captured by a secondary property is greater, equal, or smaller than the total one of the sample, the corresponding horizontal line would be above, at the same level, and below the blue one, respectively. We find that concentration, spin, and formation time capture $37\pm 2 \%$, $29\pm 2 \%$, and $25\pm 2 \%$ of the total galaxy assembly bias of this sample, respectively. We note our findings are compatible with results from semi-analytic models \citep{xuDissectingModellingGalaxy2021a}.

For a more complete picture, in Fig.~\ref{fig_GAB}, we display the redshift evolution of galaxy assembly bias for the stellar mass, $r$-band, blue, and red galaxy samples (rows) with four distinct number densities (columns). The datapoints and error bars correspond to the horizontal lines and shaded areas of Fig.~\ref{fig_GAB_instantaneous}, respectively. The top panel of the second column displays the results for the stellar mass sample with a number density of $n = 0.003\ \hMpcden$, where the values at $z = 0$ correspond to the horizontal lines in Fig.~\ref{fig_GAB_instantaneous}. For this sample, the strength of galaxy assembly bias increases as redshift decreases, increasing from 1\% at $z = 2$ to 17\% at $z = 0$. To facilitate a comparison across samples and number densities, we overlay the results for this particular sample in all other panels.

As we can see, the redshift evolution of galaxy assembly bias exhibits two primary trends, which depend on both the nature of the sample and its number density. The first trend is a steady increase in assembly bias with decreasing redshift, observed for the stellar mass and red galaxy samples with number densities greater or equal to $n = 0.001\ \hMpcden$. The second trend is a decline in assembly bias from $z=1$ to 0, found for the remaining samples and number densities. As a result of this trend, the blue galaxy sample has negative assembly bias for all number densities at $z=0$, reaching -24\% for the smallest number density considered. Consequently, the strength of galaxy assembly bias depends strongly on the properties of the target galaxy population and can increase or decrease galaxy clustering up to 25\%.

We can readily see that none of the secondary halo properties examined fully captures galaxy assembly bias for all galaxy samples. Interestingly, concentration and spin result in positive assembly bias in practically all cases, while formation time yields negative assembly bias for the r-band and blue galaxy samples. The correlation between galaxy clustering and halo formation time is therefore the most likely explanation of the negative assembly bias found for the blue galaxy sample.

\section{Predicting galaxy assembly bias}\label{sec_GAB_predict}

\begin{figure*}
    \centering
    \includegraphics[width=\textwidth]{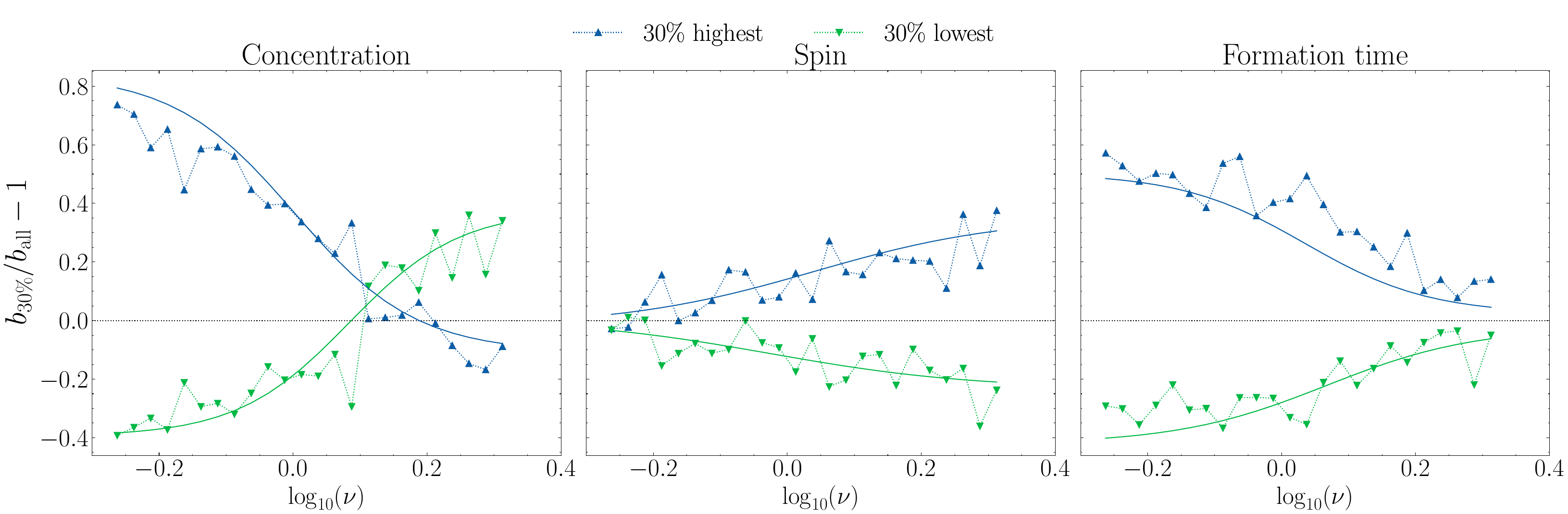}
    \caption{
    Halo assembly bias for concentration (left), spin (center), and formation time (right). In each panel, blue and green triangles display halo assembly bias for the 30\% of halos with the highest and lowest value of the corresponding property, respectively. Solid lines show the best-fitting model to the measurements (see text).
    }
    \label{fig_HAB}
\end{figure*}

In this section, we study halo assembly bias and occupancy variation in Sect. \ref{sec:GABpred_hab} and \ref{sec:GABpred_ov}, respectively, and how galaxy assembly bias emerges from their interplay in Sect. \ref{sec:GABpred_estimate}.

\subsection{Halo assembly bias}\label{sec:GABpred_hab}

The dependence of halo clustering on properties beyond halo mass is commonly known as halo assembly bias \citep[e.g.,][]{gaoAgeDependenceHalo2005a, wechslerDependenceHaloClustering2006}. Given that galaxy assembly bias refers to the dependence of galaxy clustering on halo properties beyond halo mass, halo assembly bias is a prerequisite for galaxy assembly bias. In the previous section, we computed the fraction of galaxy assembly bias attributable to concentration, spin, and formation time for multiple samples; in this section, we quantify the strength of halo assembly bias for these properties.

Halo assembly bias has no redshift dependence beyond that captured by the evolution of the density field \citep[e.g.,][]{wechslerDependenceHaloClustering2006, zentnerExcuersionSetTheory2007, gaoAssemblyBiasClustering2007}. As a result, it is useful to combine measurements of this effect from multiple redshifts to reduce uncertainties owing to the limited volume of the TNG300 simulation. To do so, we start by computing the peak height corresponding to each halo of the snapshots at $z=0$, 0.5, 1, and 2, $\nu = \delta_c / \sigma(\Mh)$, where $\delta_c = 1.686$ is the linear overdensity threshold for collapse \citep{gunnInfallMatterClusters1972} and $\sigma(\Mh)$ is the variance of the density field. We carry out this calculation using the publicly available package \texttt{colossus}\footnote{\url{https://bdiemer.bitbucket.io/colossus/}}. Then we split the halos from each snapshot into logarithmic bins of $\nu$ and select samples containing 10, 20, 30, 40, 50, and 60\% of the halos with the highest and lowest values of each secondary property from each bin. After that, we computed the clustering of sub-samples with more than 3\,000 halos, as well as the clustering of all halos within each bin. We continue by computing the average ratio of the clustering of the subsamples and all halos within each bin from $r=8$ to $25\,\hMpc$. Finally, when we have data from multiple redshifts for a particular $\nu$ bin, we compute the average of the results from all these redshifts weighted by the number of halos.

In the left, central, and right panels of Fig.~\ref{fig_HAB}, we display the level of halo assembly bias for the subsamples with the 30\% highest and lowest concentration, spin, and formation time, respectively. There is no halo assembly bias for a subsample when its clustering is the same as that of all halos, which is indicated by the dotted black line. As we can see, the dependence of halo assembly bias on halo mass is different for all properties. Low-mass halos with higher concentration are more clustered than their less concentrated counterparts, while the trend reverses for high-mass halos with peak height larger than $\log_{10} \nu \simeq 0.1$. On the other hand, halos with larger spin and forming at an earlier time are more clustered.

In Sect. \ref{sec:GABpred_estimate}, we use halo assembly bias measurements to predict galaxy assembly bias. In order to mitigate the impact of statistical noise on these predictions, we approximate HAB measurements using the following empirically motivated expression
\begin{equation}
    \mathrm{HAB}(\nu, \tilde{x}) = \frac{A(\tilde{x})}{1 + 10^{w(\tilde{x})(\nu- \nu_0(\tilde{x}))}} + B(\tilde{x}),
    \label{eq_HAB}
\end{equation}
where $A$, $B$, $w$, and $\nu_0$ are the free parameters of the model. For a subsample at a $\nu$ bin, $\tilde{x}$ indicates the ratio between the median value of a particular halo property for this subsample and the median value for all halos. In Fig.~\ref{fig_HAB}, solid lines display the predictions of the best-fitting model; as we can see, this model describes the data accurately. To remove the upturn of the spin halo assembly bias, we have to remove the backsplash halos and the halos with masses lower than $10^{10.75} \Msun$, suggesting that the effect is also due to the resolution, in addition to the backsplash halos \citep{tucci2021spinbias}.

\subsection{Occupancy Variation} 
\label{sec:GABpred_ov}

\begin{figure}
    \centering
    \includegraphics[width=\linewidth]{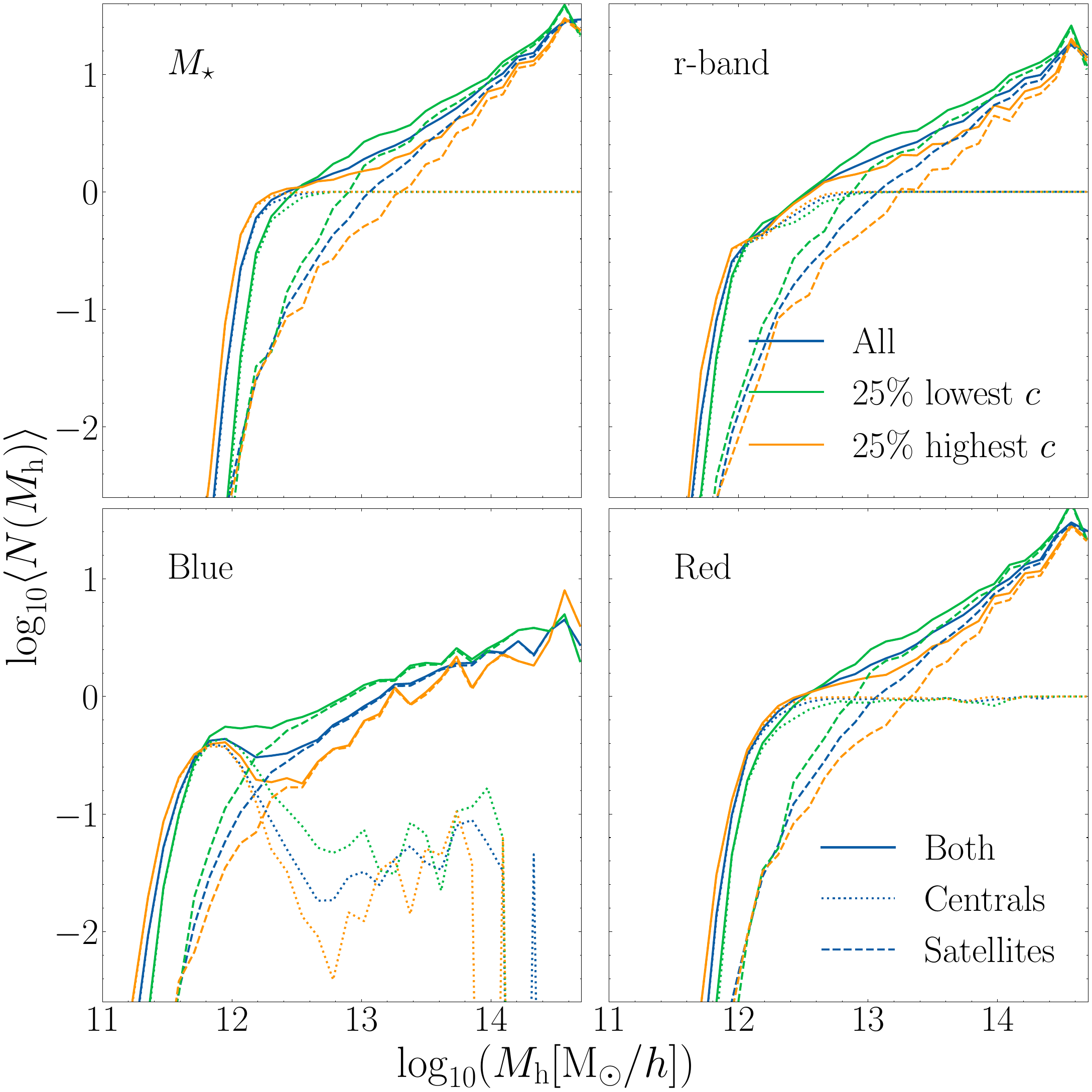}
    \caption{
    In clockwise direction, halo occupation distribution for stellar mass, r-band, blue, and red galaxy samples with $n=0.003\ \hMpcden$ at $z=0$. The solid, dotted, and dashed lines show the results for all galaxies, centrals, and satellites, respectively. The orange and green colors show the results for the 25\% of galaxies with the highest and lowest concentration, respectively, while the blue color does so for all galaxies. As we can see, the occupancy of halos depends on concentration for all galaxy samples.    
    }
    \label{fig_HOD}
\end{figure}

\begin{figure}
    \centering
    \includegraphics[width=\linewidth]{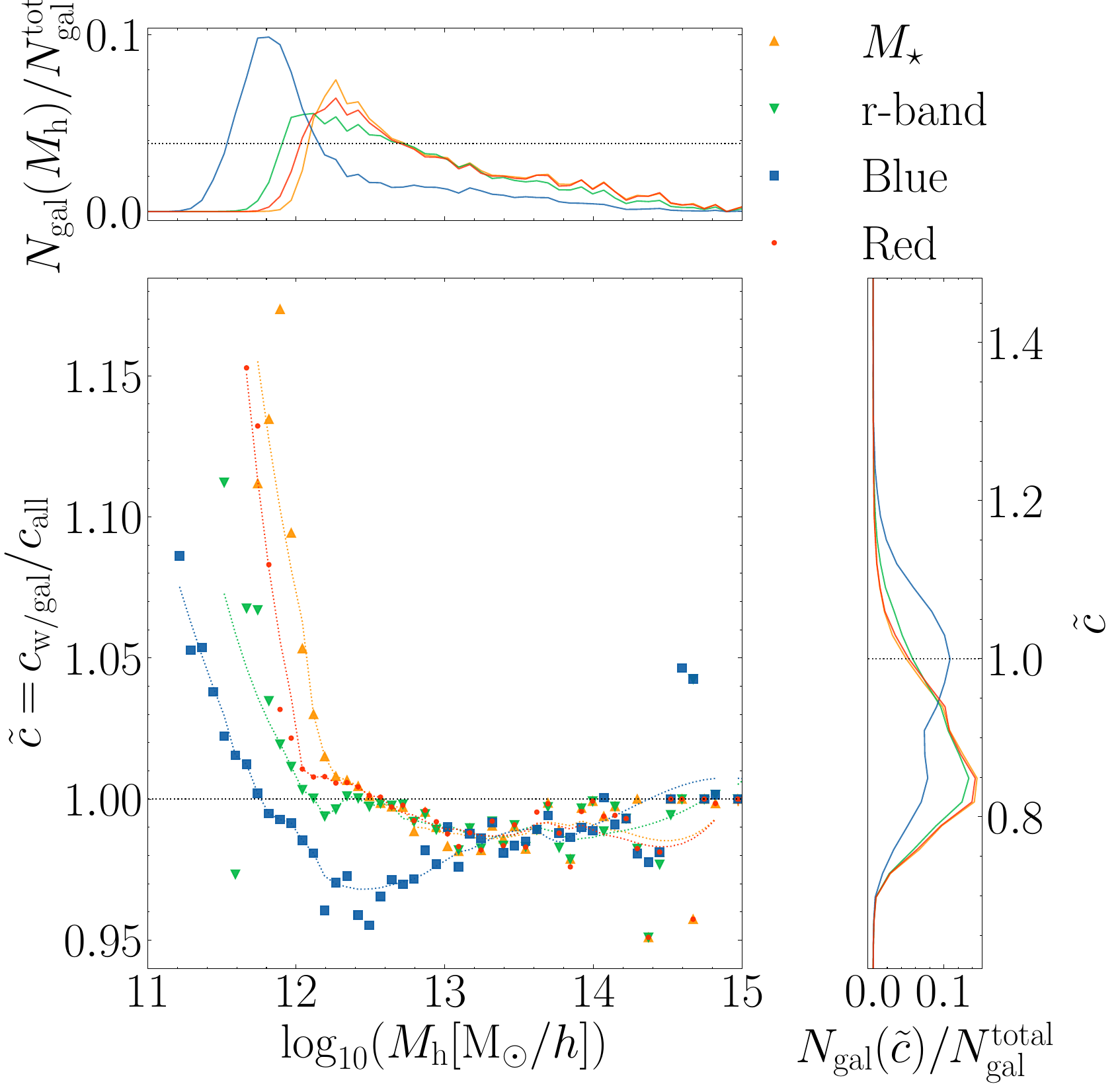}
    \caption{
    Ratio of the median concentration of the halos hosting the stellar mass, r-band, blue and red galaxy samples with $n = 0.003\ \hMpcden$ at $z=0$ and that of all halos at $z=0$. The orange, green, blue, and red colors show the results for the stellar mass, r-band, blue and red galaxy samples, respectively. The upper panel shows a histogram with the number of galaxies as a function of halo mass, while the right panel does so for the median concentration of halos hosting different galaxy populations and all halos. 
    }
    \label{fig_Feature_vs_Mhalo_0_0.003_C}
\end{figure}

Halo assembly bias is a necessary condition for galaxy assembly bias, but not sufficient. If we assume that the galaxy content of halos is independent of any property besides halo mass, galaxies would equally populate halos with values of the target property larger and smaller than average. As a result, the net halo assembly bias of the host halos could cancel out, resulting in no galaxy assembly bias for the galaxy population. On the other hand, if the galaxies of a particular sample preferentially populate, for example, older halos, we would expect positive assembly bias given that the halo assembly bias for these halos is positive.

The dependence of the galaxy content of halos on secondary halo properties beyond halo mass is commonly known as occupancy variation, and it is common to study this effect by computing the occupation distribution of halos with different values of secondary properties \citep[e.g.,][]{zehaviImpactAssemblyBias2018a, artaleImpactAssemblyBias2018a, contrerasEvolutionAssemblyBias2019a}. In the top-left, top-right, bottom-left, and bottom-right panels of Fig.~\ref{fig_HOD}, blue lines display the average halo occupation for the stellar mass, r-band, blue, and red samples with $n = 0.003\ \hMpcden$ at $z=0$, respectively, while the green and orange lines show the halo occupation of the halos with the 30\% highest and lowest concentration. For all samples, the probability of hosting satellites increases with halo mass, while the probability of hosting centrals increases with halo mass for all samples but the blue one, which decreases after peaking for small halo masses. This trend is explained by the fact that more massive halos typically host redder galaxies \citep[e.g.,][]{chaves-montero2020SurrogateModellingBaryonic}. 

More interestingly, we find that low-mass halos with high concentration are more likely to host a central galaxy than those with low concentration for all samples, while the trend reverses for satellite galaxies and centrals in high-mass halos of the blue sample. Nevertheless, these dependencies are not very informative for the level of galaxy assembly bias of the sample. This is because galaxies could occupy halos with a mean value of a secondary property equal to that of all halos in the simulation; as a result, the halo assembly bias of the halos hosting galaxies would be zero. Furthermore, if the occupation content of these halos is symmetric for values of the target property higher and lower than the mean, the galaxy assembly bias of the sample could also be zero.

To better study the impact of occupancy variation on galaxy assembly bias, in Fig.~\ref{fig_Feature_vs_Mhalo_0_0.003_C} we display the ratio between the median concentration of halos hosting the galaxies of the samples shown in Fig.~\ref{fig_HOD} and all halos at this redshift, $\tilde{c}$. The value of $\tilde{c}$ is greater (smaller) than unity if galaxies preferentially populate more (less) concentrated halos. The stellar mass, r-band, and red samples prefer to occupy more (less) concentrated halos with masses smaller (larger) than $\Mh = 10^{12.6}\ \hMsun$. At $z=0$, halo assembly bias is positive for more (less) concentrated halos with masses smaller (larger) than $\Mh = 10^{13.4}\ \hMsun$. As a result, the majority of the galaxies of these samples populate halos with a positive halo assembly bias, which explains the positive galaxy assembly bias of these samples. On the other hand, blue galaxies populate more (less) concentrated halos that are less (more) massive than $\Mh = 10^{11.75}\ \hMsun$. Consequently, the majority of the galaxies of this sample populate halos with negative halo assembly bias; as a result, the galaxy assembly bias of this sample is negative (see Fig.~\ref{fig_GAB}). 

In the next section, we use occupancy variation measurements to predict galaxy assembly bias. In order to mitigate the impact of statistical noise, we approximate occupancy variation measurements using a Savitzky–Golay filter when the number of galaxies in a halo mass bin is smaller than $1000$. Dotted lines display the resulting curves.

\subsection{Predicting galaxy assembly bias}\label{sec:GABpred_estimate}

\begin{figure}
    \centering
    \includegraphics[width=\linewidth]{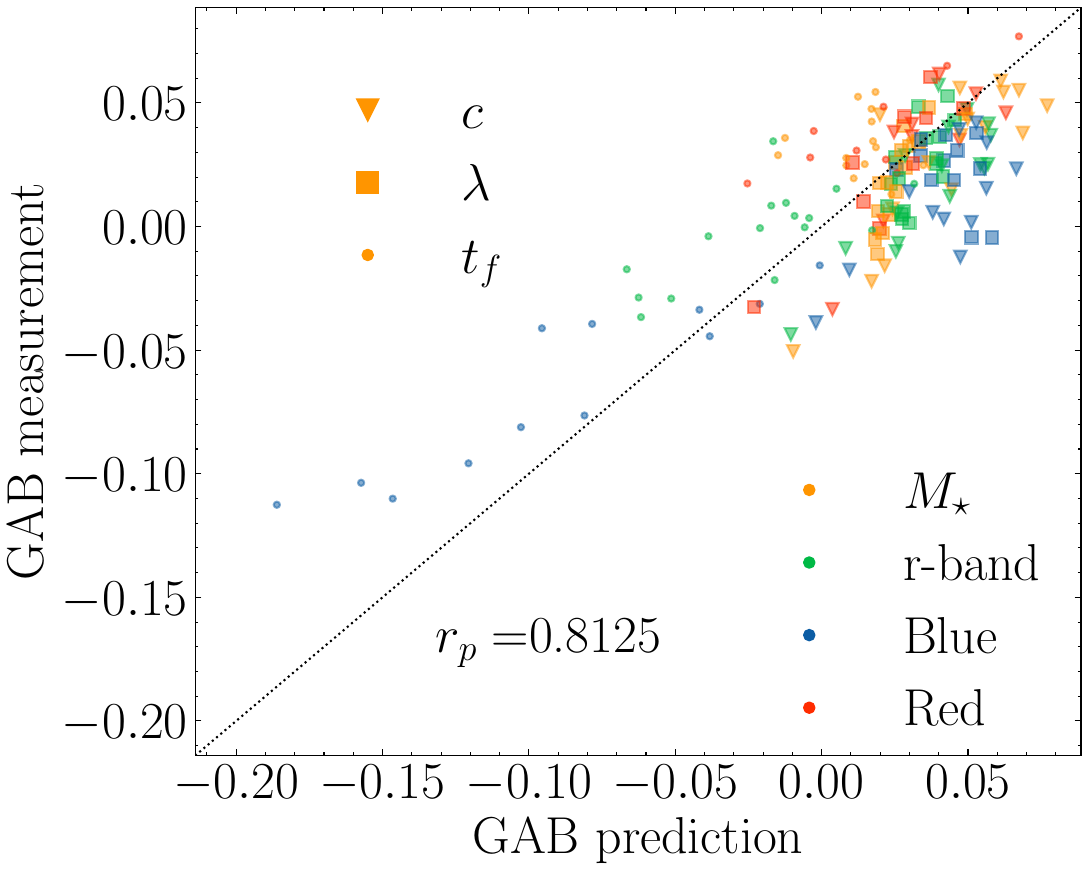}
    \caption{
    Comparison between direct measurements of galaxy assembly bias from the shuffling procedure and predictions from our analytic model. Orange, green, blue, and red colors show the results for the stellar mass, $r$-band, blue, and red samples, respectively, while the triangles, squares and dots indicate the galaxy assembly bias due to concentration, spin, and formation time. The dotted line indicates a $1:1$ relation between measurements and predictions. We can readily see that our analytic expression accurately predicts the amount of galaxy assembly bias resulting from the three studied halo properties, with a Pearson correlation greater than $r=0.8$.
    }
    \label{fig_GAB_Correlation}
\end{figure}

As discussed in the previous two sections, galaxy assembly bias emerges from the interplay of halo assembly bias and occupancy variation. In this section, we provide an analytic expression to efficiently predict galaxy assembly bias as a function of these two effects without the need to employ computationally expensive shuffling techniques.

In the absence of galaxy assembly bias, the linear bias of a galaxy sample can be expressed as follows
\begin{equation}
    \label{eq:bnoHAB}
    b_{\rm gal}^{\rm w/o\ HAB} = \frac{\int \mathrm{d}\Mh b_\mathrm{h}(\Mh) N_{\rm gal}(\Mh)}{\int \mathrm{d}\Mh N_{\rm gal}(\Mh)},
\end{equation}
where $b_\mathrm{h}(\Mh)$ is the linear bias of a halo of mass $\Mh$. We compute the halo bias using the \citet{comparatAccurateMassVelocity2017} parameterization as implemented in \texttt{colossus}. We can modify the previous expression as follows to account for the galaxy assembly bias induced by a halo property $x$
\begin{equation}
    \label{eq:bHAB}
    b_{\rm gal}^{\rm w/\ HAB}(\tilde{x}) = \frac{\int \mathrm{d}\Mh b_\mathrm{h}(\Mh)[1 + {\rm HAB}(\Mh, \tilde{x}(\Mh))] N_{\rm gal}(\Mh)}{\int \mathrm{d}\Mh N_{\rm gal}(\Mh)},
\end{equation}
where $\tilde{x}(\Mh)$ is the ratio between the median value of the target halo property for halos containing galaxies and that of all halos as a function of halo mass (see Sect. \ref{sec:GABpred_ov}) and ${\rm HAB}(\Mh, \tilde{x}(\Mh))$ is the strength of halo assembly bias as a function of halo mass and $\tilde{x}(\Mh)$. Note that this expression deviates from eq.~17 of \citet{wechslerDependenceHaloClustering2006} since it does not require computing the distribution of galaxies and linear halo bias as a function of halo mass and property $x$, nor the probability of finding a halo of property $x$ as a function of halo mass.

Combining the previous two equations allows us to compute the galaxy assembly bias induced by a halo property $x$,
\begin{equation}
    \label{eq:pred_gab}
    \mathrm{GAB}(x) = b_{\rm gal}^{\rm w/\,HAB}(\tilde{x})/b_{\rm gal}^{\rm w/o\,HAB},
\end{equation}
where this expression provides a rapid estimate of galaxy assembly bias without relying on computationally expensive shuffling techniques. In order to improve the accuracy of this expression, we could measure the level of halo assembly bias for different halo properties from any large gravity-only simulation, given that galaxy assembly bias is largely insensitive to cosmology \citep{contrerasEvolutionAssemblyBias2019a}.

In Fig.~\ref{fig_GAB_Correlation}, we compare measurements of the galaxy assembly bias induced by halo concentration, spin, and formation time obtained from the shuffling technique with predictions from the above expression. We show this comparison for all the samples, number densities, and redshifts analyzed in this work. As shown, Eq.~\ref{eq:pred_gab} accurately reproduces the shuffling measurements for both positive and negative galaxy assembly bias, yielding a Pearson correlation coefficient of $r_p = 0.8$. We therefore conclude that Eq.~\ref{eq:pred_gab} provides an accurate prediction of the galaxy assembly bias produced by a given halo property.

\section{Summary and conclusions} \label{sec_Conclusions}

In this work, we studied the level of galaxy assembly bias for galaxy populations selected according to stellar mass, luminosity, and broad-band colors from the largest hydrodynamical simulation from the IllustrisTNG suite. We summarize our main findings below:

\begin{itemize}
    \item In Fig.~\ref{fig_GAB}, we show that the strength of galaxy assembly bias depends strongly upon the selection criteria, number density, and redshift of the galaxy sample, increasing or decreasing clustering by as much as 25\%. 
    
    \item The amount of galaxy assembly bias emerging from halo concentration, spin, and formation time does not fully explain the strength of this effect for any galaxy sample studied in this work. As a result, empirical approaches modeling galaxy assembly bias as a function of halo concentration or other single halo property cannot reproduce predictions from hydrodynamical simulations.
    
    \item In Sect. \ref{sec_GAB_predict}, we provide an analytical expression to estimate the galaxy assembly bias induced by a halo property without having to resort to computationally expensive shuffling techniques. We validate that this expression accurately reproduces the level of galaxy assembly bias estimated by shuffling procedures, with a Pearson correlation coefficient greater than $r = 0.8$.
\end{itemize}

This work conducts a systematic analysis of galaxy assembly bias for galaxy samples with characteristics similar to those used for cosmological inference in state-of-the-art spectroscopic and photometric surveys. We find that the impact of galaxy assembly bias on galaxy clustering cannot be neglected for any galaxy sample, and thus it is crucial to account for this effect in the modeling of nonlinear scales for unbiased cosmological inference \citep[e.g.,][]{chavesGalFormationOriginLensingLow2023, contrerasbosslil2023}.

To facilitate the modeling of galaxy assembly bias, we provide a fast analytical expression that predicts the level of galaxy assembly bias induced by any halo property, eliminating the need for computationally intensive shuffling techniques. This expression requires as input the level of halo assembly bias associated with a given halo property ---measurable from any cosmological simulation --- and the dependence of galaxy occupation on that property, which can be obtained from any galaxy-halo connection model. We envision using this framework to rapidly assess whether the level of galaxy assembly bias introduced by an empirical model is consistent with predictions from hydrodynamical simulations and semi-analytic models.

\begin{acknowledgements}
    We thank the anonymous referee for the thorough review, insightful comments, and positive suggestions. We thank Sergio Contreras for useful comments and discussion, as well as the IllustrisTNG collaboration for making their data publicly available. JCM acknowledges support from the European Union’s Horizon Europe research and innovation programme (COSMO-LYA, grant agreement 101044612). We acknowledge the use of the Port d’Informació Científica (PIC) for the analysis, which is maintained through a collaboration agreement between the Institut de Física d’Altes Energies (IFAE) and the Centro de Investigaciones Energéticas, Medioambientales y Tecnológicas (CIEMAT). IFAE is partially funded by the CERCA program of the Generalitat de Catalunya. This work made direct use of the following software Python packages: \texttt{Matplotlib} \citep{hunterMatplotlib2DGraphics2007}, \verb|collossus| \citep{diemerCOLOSSUSPythonToolkit2018}, \verb|Corrfunc| \citep{sinhaCorrfuncSuiteBlazing2020a}, \verb|Numpy| \citep{harrisArrayProgrammingNumPy2020}, \verb|Scipy| \citep{virtanenSciPy10Fundamental2020}, \verb|Seaborn| \citep{waskomSeabornStatisticalData2021}, \verb|SciencPlots| \citep{garrettSciencePlots2112023} and \verb|pandas| \citep{mckinney-proc-scipy-2010, thepandasdevelopmentteamPandas2024}.
\end{acknowledgements}

\bibliographystyle{aa_url}
\bibliography{aa}

\begin{thebibliography}{64}
\expandafter\ifx\csname natexlab\endcsname\relax\def\natexlab#1{#1}\fi

\bibitem[{{Alam} {et~al.}(2024){Alam}, {Paranjape}, \&
  {Peacock}}]{Alam2024_gab}
{Alam}, S., {Paranjape}, A., \& {Peacock}, J.~A. 2024,
  \href{http://dx.doi.org/10.1093/mnras/stad3423}{\color{magenta}\mnras},
  \href{https://ui.adsabs.harvard.edu/abs/2024MNRAS.527.3771A}{527, 3771}

\bibitem[{{Artale} {et~al.}(2018){Artale}, {Zehavi}, {Contreras}, \&
  {Norberg}}]{artaleImpactAssemblyBias2018a}
{Artale}, M.~C., {Zehavi}, I., {Contreras}, S., \& {Norberg}, P. 2018,
  \href{http://dx.doi.org/10.1093/mnras/sty2110}{\color{magenta}\mnras},
  \href{https://ui.adsabs.harvard.edu/abs/2018MNRAS.480.3978A}{480, 3978}

\bibitem[{{Asgari} {et~al.}(2021){Asgari}, {Lin}, {Joachimi}, {Giblin},
  {Heymans}, {Hildebrandt}, {Kannawadi}, {St{\"o}lzner}, {Tr{\"o}ster}, {van
  den Busch}, {Wright}, {Bilicki}, {Blake}, {de Jong}, {Dvornik}, {Erben},
  {Getman}, {Hoekstra}, {K{\"o}hlinger}, {Kuijken}, {Miller}, {Radovich},
  {Schneider}, {Shan}, \& {Valentijn}}]{kids2021}
{Asgari}, M., {Lin}, C.-A., {Joachimi}, B., {et~al.} 2021,
  \href{http://dx.doi.org/10.1051/0004-6361/202039070}{\color{magenta}\aap},
  \href{https://ui.adsabs.harvard.edu/abs/2021A&A...645A.104A}{645, A104}

\bibitem[{{Bullock} {et~al.}(2001){Bullock}, {Dekel}, {Kolatt}, {Kravtsov},
  {Klypin}, {Porciani}, \& {Primack}}]{bullockUniversalAngularMomentum2001a}
{Bullock}, J.~S., {Dekel}, A., {Kolatt}, T.~S., {et~al.} 2001,
  \href{http://dx.doi.org/10.1086/321477}{\color{magenta}\apj},
  \href{https://ui.adsabs.harvard.edu/abs/2001ApJ...555..240B}{555, 240}

\bibitem[{{Chaves-Montero} {et~al.}(2023){Chaves-Montero}, {Angulo}, \&
  {Contreras}}]{chavesGalFormationOriginLensingLow2023}
{Chaves-Montero}, J., {Angulo}, R.~E., \& {Contreras}, S. 2023,
  \href{http://dx.doi.org/10.1093/mnras/stad243}{\color{magenta}\mnras},
  \href{https://ui.adsabs.harvard.edu/abs/2023MNRAS.521..937C}{521, 937}

\bibitem[{{Chaves-Montero} {et~al.}(2016){Chaves-Montero}, {Angulo}, {Schaye},
  {Schaller}, {Crain}, {Furlong}, \&
  {Theuns}}]{chaves-monteroSubhaloAbundanceMatching2016}
{Chaves-Montero}, J., {Angulo}, R.~E., {Schaye}, J., {et~al.} 2016,
  \href{http://dx.doi.org/10.1093/mnras/stw1225}{\color{magenta}\mnras},
  \href{https://ui.adsabs.harvard.edu/abs/2016MNRAS.460.3100C}{460, 3100}

\bibitem[{{Chaves-Montero} \&
  Hearin(2020)}]{chaves-montero2020SurrogateModellingBaryonic}
{Chaves-Montero}, J. \& Hearin, A. 2020,
  \href{http://dx.doi.org/10.1093/mnras/staa1230}{\color{magenta}\mnras},
  \href{http://adsabs.harvard.edu/abs/2020MNRAS.495.2088C}{495, 2088}

\bibitem[{{Child} {et~al.}(2018){Child}, {Habib}, {Heitmann}, {Frontiere},
  {Finkel}, {Pope}, \& {Morozov}}]{childHaloProfilesConcentration2018}
{Child}, H.~L., {Habib}, S., {Heitmann}, K., {et~al.} 2018,
  \href{http://dx.doi.org/10.3847/1538-4357/aabf95}{\color{magenta}\apj},
  \href{https://ui.adsabs.harvard.edu/abs/2018ApJ...859...55C}{859, 55}

\bibitem[{{Comparat} {et~al.}(2017){Comparat}, {Prada}, {Yepes}, \&
  {Klypin}}]{comparatAccurateMassVelocity2017}
{Comparat}, J., {Prada}, F., {Yepes}, G., \& {Klypin}, A. 2017,
  \href{http://dx.doi.org/10.1093/mnras/stx1183}{\color{magenta}\mnras},
  \href{https://ui.adsabs.harvard.edu/abs/2017MNRAS.469.4157C}{469, 4157}

\bibitem[{{Contreras} {et~al.}(2023{\natexlab{a}}){Contreras}, {Angulo},
  {Springel}, {White}, {Hadzhiyska}, {Hernquist}, {Pakmor}, {Kannan},
  {Hern{\'a}ndez-Aguayo}, {Barrera}, {Ferlito}, {Delgado}, {Bose}, \&
  {Frenk}}]{contreras2023MilleniumTNG}
{Contreras}, S., {Angulo}, R.~E., {Springel}, V., {et~al.} 2023{\natexlab{a}},
  \href{http://dx.doi.org/10.1093/mnras/stac3699}{\color{magenta}\mnras},
  \href{https://ui.adsabs.harvard.edu/abs/2023MNRAS.524.2489C}{524, 2489}

\bibitem[{{Contreras} {et~al.}(2021){Contreras}, {Angulo}, \&
  {Zennaro}}]{contreras2021FlexibleModellingGalaxy}
{Contreras}, S., {Angulo}, R.~E., \& {Zennaro}, M. 2021,
  \href{http://dx.doi.org/10.1093/mnras/stab1170}{\color{magenta}\mnras},
  \href{https://ui.adsabs.harvard.edu/abs/2021MNRAS.504.5205C}{504, 5205}

\bibitem[{{Contreras} {et~al.}(2023{\natexlab{b}}){Contreras},
  {Chaves-Montero}, \& {Angulo}}]{contrerasbosslil2023}
{Contreras}, S., {Chaves-Montero}, J., \& {Angulo}, R.~E. 2023{\natexlab{b}},
  \href{http://dx.doi.org/10.1093/mnras/stad2434}{\color{magenta}\mnras},
  \href{https://ui.adsabs.harvard.edu/abs/2023MNRAS.525.3149C}{525, 3149}

\bibitem[{{Contreras} {et~al.}(2019){Contreras}, {Zehavi}, {Padilla}, {Baugh},
  {Jim{\'e}nez}, \& {Lacerna}}]{contrerasEvolutionAssemblyBias2019a}
{Contreras}, S., {Zehavi}, I., {Padilla}, N., {et~al.} 2019,
  \href{http://dx.doi.org/10.1093/mnras/stz018}{\color{magenta}\mnras},
  \href{https://ui.adsabs.harvard.edu/abs/2019MNRAS.484.1133C}{484, 1133}

\bibitem[{{Croton} {et~al.}(2007){Croton}, {Gao}, \&
  {White}}]{crotonHaloAssemblyBias2007a}
{Croton}, D.~J., {Gao}, L., \& {White}, S. D.~M. 2007,
  \href{http://dx.doi.org/10.1111/j.1365-2966.2006.11230.x}{\color{magenta}\mnras},
  \href{https://ui.adsabs.harvard.edu/abs/2007MNRAS.374.1303C}{374, 1303}

\bibitem[{{Davis} {et~al.}(1985){Davis}, {Efstathiou}, {Frenk}, \&
  {White}}]{davisEvolutionLargescaleStructure1985}
{Davis}, M., {Efstathiou}, G., {Frenk}, C.~S., \& {White}, S.~D.~M. 1985,
  \href{http://dx.doi.org/10.1086/163168}{\color{magenta}\apj},
  \href{https://ui.adsabs.harvard.edu/abs/1985ApJ...292..371D}{292, 371}

\bibitem[{{DESI Collaboration} {et~al.}(2016){DESI Collaboration}, {Aghamousa},
  {Aguilar}, {Ahlen}, {Alam}, {Allen}, {Allende Prieto}, {Annis}, {Bailey},
  {Balland}, {Ballester}, {Baltay}, {Beaufore}, {Bebek}, {Beers}, {Bell},
  {Bernal}, {Besuner}, {Beutler}, {Blake}, {Bleuler}, {Blomqvist}, {Blum},
  {Bolton}, {Briceno}, {Brooks}, {Brownstein}, {Buckley-Geer}, {Burden},
  {Burtin}, {Busca}, {Cahn}, {Cai}, {Cardiel-Sas}, {Carlberg}, {Carton},
  {Casas}, {Castander}, {Cervantes-Cota}, {Claybaugh}, {Close}, {Coker},
  {Cole}, {Comparat}, {Cooper}, {Cousinou}, {Crocce}, {Cuby}, {Cunningham},
  {Davis}, {Dawson}, {de la Macorra}, {De Vicente}, {Delubac}, {Derwent},
  {Dey}, {Dhungana}, {Ding}, {Doel}, {Duan}, {Ealet}, {Edelstein},
  {Eftekharzadeh}, {Eisenstein}, {Elliott}, {Escoffier}, {Evatt}, {Fagrelius},
  {Fan}, {Fanning}, {Farahi}, {Farihi}, {Favole}, {Feng}, {Fernandez},
  {Findlay}, {Finkbeiner}, {Fitzpatrick}, {Flaugher}, {Flender}, {Font-Ribera},
  {Forero-Romero}, {Fosalba}, {Frenk}, {Fumagalli}, {Gaensicke}, {Gallo},
  {Garcia-Bellido}, {Gaztanaga}, {Pietro Gentile Fusillo}, {Gerard},
  {Gershkovich}, {Giannantonio}, {Gillet}, {Gonzalez-de-Rivera},
  {Gonzalez-Perez}, {Gott}, {Graur}, {Gutierrez}, {Guy}, {Habib}, {Heetderks},
  {Heetderks}, {Heitmann}, {Hellwing}, {Herrera}, {Ho}, {Holland}, {Honscheid},
  {Huff}, {Hutchinson}, {Huterer}, {Hwang}, {Illa Laguna}, {Ishikawa},
  {Jacobs}, {Jeffrey}, {Jelinsky}, {Jennings}, {Jiang}, {Jimenez}, {Johnson},
  {Joyce}, {Jullo}, {Juneau}, {Kama}, {Karcher}, {Karkar}, {Kehoe}, {Kennamer},
  {Kent}, {Kilbinger}, {Kim}, {Kirkby}, {Kisner}, {Kitanidis}, {Kneib},
  {Koposov}, {Kovacs}, {Koyama}, {Kremin}, {Kron}, {Kronig}, {Kueter-Young},
  {Lacey}, {Lafever}, {Lahav}, {Lambert}, {Lampton}, {Landriau}, {Lang},
  {Lauer}, {Le Goff}, {Le Guillou}, {Le Van Suu}, {Lee}, {Lee}, {Leitner},
  {Lesser}, {Levi}, {L'Huillier}, {Li}, {Liang}, {Lin}, {Linder}, {Loebman},
  {Luki{\'c}}, {Ma}, {MacCrann}, {Magneville}, {Makarem}, {Manera}, {Manser},
  {Marshall}, {Martini}, {Massey}, {Matheson}, {McCauley}, {McDonald},
  {McGreer}, {Meisner}, {Metcalfe}, {Miller}, {Miquel}, {Moustakas}, {Myers},
  {Naik}, {Newman}, {Nichol}, {Nicola}, {Nicolati da Costa}, {Nie}, {Niz},
  {Norberg}, {Nord}, {Norman}, {Nugent}, {O'Brien}, {Oh}, \&
  {Olsen}}]{DESI2016a.Science}
{DESI Collaboration}, {Aghamousa}, A., {Aguilar}, J., {et~al.} 2016,
  \href{https://ui.adsabs.harvard.edu/abs/2016arXiv161100036D}{\href{http://dx.doi.org/10.48550/arXiv.1611.00036}{\color{magenta}arXiv
  e-prints}, arXiv:1611.00036}

\bibitem[{{Diemer}(2018)}]{diemerCOLOSSUSPythonToolkit2018}
{Diemer}, B. 2018,
  \href{http://dx.doi.org/10.3847/1538-4365/aaee8c}{\color{magenta}\apjs},
  \href{https://ui.adsabs.harvard.edu/abs/2018ApJS..239...35D}{239, 35}

\bibitem[{{Dolag} {et~al.}(2009){Dolag}, {Borgani}, {Murante}, \&
  {Springel}}]{dolagSubstructuresHydrodynamicalCluster2009}
{Dolag}, K., {Borgani}, S., {Murante}, G., \& {Springel}, V. 2009,
  \href{http://dx.doi.org/10.1111/j.1365-2966.2009.15034.x}{\color{magenta}\mnras},
  \href{https://ui.adsabs.harvard.edu/abs/2009MNRAS.399..497D}{399, 497}

\bibitem[{{Gao} {et~al.}(2005){Gao}, {Springel}, \&
  {White}}]{gaoAgeDependenceHalo2005a}
{Gao}, L., {Springel}, V., \& {White}, S. D.~M. 2005,
  \href{http://dx.doi.org/10.1111/j.1745-3933.2005.00084.x}{\color{magenta}\mnras},
  \href{https://ui.adsabs.harvard.edu/abs/2005MNRAS.363L..66G}{363, L66}

\bibitem[{{Gao} \& {White}(2007)}]{gaoAssemblyBiasClustering2007}
{Gao}, L. \& {White}, S. D.~M. 2007,
  \href{http://dx.doi.org/10.1111/j.1745-3933.2007.00292.x}{\color{magenta}\mnras},
  \href{https://ui.adsabs.harvard.edu/abs/2007MNRAS.377L...5G}{377, L5}

\bibitem[{{Garrett} {et~al.}(2023){Garrett}, {Luis}, {Peng}, {Cera},
  {gobinathj}, {Borrow}, {Ke{\c{c}}eci}, {Splines}, {Iyer}, {Liu}, {cjw}, \&
  {Gasanov}}]{garrettSciencePlots2112023}
{Garrett}, J., {Luis}, E., {Peng}, H.-H., {et~al.} 2023,
  {garrettj403/SciencePlots: 2.1.1}

\bibitem[{{Gunn} \& {Gott}(1972)}]{gunnInfallMatterClusters1972}
{Gunn}, J.~E. \& {Gott}, III, J.~R. 1972,
  \href{http://dx.doi.org/10.1086/151605}{\color{magenta}\apj},
  \href{https://ui.adsabs.harvard.edu/abs/1972ApJ...176....1G}{176, 1}

\bibitem[{{Hadzhiyska} {et~al.}(2021{\natexlab{a}}){Hadzhiyska}, {Liu},
  {Somerville}, {Gabrielpillai}, {Bose}, {Eisenstein}, \&
  {Hernquist}}]{hadzhiyska2021GalaxyBiasIllustrisTNGSantaCruz}
{Hadzhiyska}, B., {Liu}, S., {Somerville}, R.~S., {et~al.} 2021{\natexlab{a}},
  \href{http://dx.doi.org/10.1093/mnras/stab2564}{\color{magenta}\mnras},
  \href{https://ui.adsabs.harvard.edu/abs/2021MNRAS.508..698H}{508, 698}

\bibitem[{{Hadzhiyska} {et~al.}(2021{\natexlab{b}}){Hadzhiyska}, {Tacchella},
  {Bose}, \& {Eisenstein}}]{hadzhiyska2021GalaxyHaloConnectionIllustrisTNG}
{Hadzhiyska}, B., {Tacchella}, S., {Bose}, S., \& {Eisenstein}, D.~J.
  2021{\natexlab{b}},
  \href{http://dx.doi.org/10.1093/mnras/stab243}{\color{magenta}\mnras},
  \href{https://ui.adsabs.harvard.edu/abs/2021MNRAS.502.3599H}{502, 3599}

\bibitem[{{Harris} {et~al.}(2020){Harris}, {Millman}, {van der Walt},
  {Gommers}, {Virtanen}, {Cournapeau}, {Wieser}, {Taylor}, {Berg}, {Smith},
  {Kern}, {Picus}, {Hoyer}, {van Kerkwijk}, {Brett}, {Haldane}, {del R{\'\i}o},
  {Wiebe}, {Peterson}, {G{\'e}rard-Marchant}, {Sheppard}, {Reddy}, {Weckesser},
  {Abbasi}, {Gohlke}, \& {Oliphant}}]{harrisArrayProgrammingNumPy2020}
{Harris}, C.~R., {Millman}, K.~J., {van der Walt}, S.~J., {et~al.} 2020,
  \href{http://dx.doi.org/10.1038/s41586-020-2649-2}{\color{magenta}\nat},
  \href{https://ui.adsabs.harvard.edu/abs/2020Natur.585..357H}{585, 357}

\bibitem[{{Hunter}(2007)}]{hunterMatplotlib2DGraphics2007}
{Hunter}, J.~D. 2007,
  \href{http://dx.doi.org/10.1109/MCSE.2007.55}{\color{magenta}CiSE},
  \href{https://ui.adsabs.harvard.edu/abs/2007CSE.....9...90H}{9, 90}

\bibitem[{{Jimenez} \& {Heavens}(2020)}]{jimenez2020DistributionDarkGalaxies}
{Jimenez}, R. \& {Heavens}, A.~F. 2020,
  \href{http://dx.doi.org/10.1093/mnrasl/slaa135}{\color{magenta}\mnras},
  \href{https://ui.adsabs.harvard.edu/abs/2020MNRAS.498L..93J}{498, L93}

\bibitem[{{Kong} {et~al.}(2025){Kong}, {Boylan-Kolchin}, \&
  {Bullock}}]{kong_bloodhound_2025}
{Kong}, H., {Boylan-Kolchin}, M., \& {Bullock}, J.~S. 2025,
  \href{https://ui.adsabs.harvard.edu/abs/2025arXiv250310766K}{\href{http://dx.doi.org/10.48550/arXiv.2503.10766}{\color{magenta}arXiv
  e-prints}, arXiv:2503.10766}

\bibitem[{{Marinacci} {et~al.}(2018){Marinacci}, {Vogelsberger}, {Pakmor},
  {Torrey}, {Springel}, {Hernquist}, {Nelson}, {Weinberger}, {Pillepich},
  {Naiman}, \& {Genel}}]{marinacciFirstResultsIllustrisTNG2018a}
{Marinacci}, F., {Vogelsberger}, M., {Pakmor}, R., {et~al.} 2018,
  \href{http://dx.doi.org/10.1093/mnras/sty2206}{\color{magenta}\mnras},
  \href{https://ui.adsabs.harvard.edu/abs/2018MNRAS.480.5113M}{480, 5113}

\bibitem[{{McCarthy} {et~al.}(2022){McCarthy}, {Zheng}, {Guo}, {Luo}, \&
  {Lin}}]{McCarthy2022_gab}
{McCarthy}, K.~S., {Zheng}, Z., {Guo}, H., {Luo}, W., \& {Lin}, Y.-T. 2022,
  \href{http://dx.doi.org/10.1093/mnras/stab2602}{\color{magenta}\mnras},
  \href{https://ui.adsabs.harvard.edu/abs/2022MNRAS.509..380M}{509, 380}

\bibitem[{{Naiman} {et~al.}(2018){Naiman}, {Pillepich}, {Springel},
  {Ramirez-Ruiz}, {Torrey}, {Vogelsberger}, {Pakmor}, {Nelson}, {Marinacci},
  {Hernquist}, {Weinberger}, \& {Genel}}]{naimanFirstResultsIllustrisTNG2018a}
{Naiman}, J.~P., {Pillepich}, A., {Springel}, V., {et~al.} 2018,
  \href{http://dx.doi.org/10.1093/mnras/sty618}{\color{magenta}\mnras},
  \href{https://ui.adsabs.harvard.edu/abs/2018MNRAS.477.1206N}{477, 1206}

\bibitem[{{Nelson} {et~al.}(2018){Nelson}, {Pillepich}, {Springel},
  {Weinberger}, {Hernquist}, {Pakmor}, {Genel}, {Torrey}, {Vogelsberger},
  {Kauffmann}, {Marinacci}, \& {Naiman}}]{nelsonFirstResultsIllustrisTNG2018a}
{Nelson}, D., {Pillepich}, A., {Springel}, V., {et~al.} 2018,
  \href{http://dx.doi.org/10.1093/mnras/stx3040}{\color{magenta}\mnras},
  \href{https://ui.adsabs.harvard.edu/abs/2018MNRAS.475..624N}{475, 624}

\bibitem[{{Niemiec} {et~al.}(2018){Niemiec}, {Jullo}, {Montero-Dorta}, {Prada},
  {Rodriguez-Torres}, {Perez}, {Klypin}, {Erben}, {Makler}, {Moraes},
  {Pereira}, \& {Shan}}]{niemiec2018GalaxyBiasLRGweakLensing}
{Niemiec}, A., {Jullo}, E., {Montero-Dorta}, A.~D., {et~al.} 2018,
  \href{http://dx.doi.org/10.1093/mnrasl/sly041}{\color{magenta}\mnras},
  \href{https://ui.adsabs.harvard.edu/abs/2018MNRAS.477L...1N}{477, L1}

\bibitem[{{Paviot} {et~al.}(2024){Paviot}, {Rocher}, {Codis}, {de Mattia},
  {Jullo}, \& {de la Torre}}]{Paviot2024_gab}
{Paviot}, R., {Rocher}, A., {Codis}, S., {et~al.} 2024,
  \href{http://dx.doi.org/10.1051/0004-6361/202449574}{\color{magenta}\aap},
  \href{https://ui.adsabs.harvard.edu/abs/2024A&A...690A.221P}{690, A221}

\bibitem[{{Pearl} {et~al.}(2024){Pearl}, {Zentner}, {Newman}, {Bezanson},
  {Wang}, {Moustakas}, {Aguilar}, {Ahlen}, {Brooks}, {Claybaugh}, {Cole},
  {Dawson}, {de la Macorra}, {Doel}, {Forero-Romero}, {Gontcho A Gontcho},
  {Honscheid}, {Landriau}, {Manera}, {Martini}, {Meisner}, {Miquel}, {Nie},
  {Percival}, {Prada}, {Rezaie}, {Rossi}, {Sanchez}, {Schubnell}, {Tarl{\'e}},
  {Weaver}, \& {Zhou}}]{Pearl2024_gab}
{Pearl}, A.~N., {Zentner}, A.~R., {Newman}, J.~A., {et~al.} 2024,
  \href{http://dx.doi.org/10.3847/1538-4357/ad1ffd}{\color{magenta}\apj},
  \href{https://ui.adsabs.harvard.edu/abs/2024ApJ...963..116P}{963, 116}

\bibitem[{{Pillepich} {et~al.}(2018{\natexlab{a}}){Pillepich}, {Nelson},
  {Hernquist}, {Springel}, {Pakmor}, {Torrey}, {Weinberger}, {Genel}, {Naiman},
  {Marinacci}, \& {Vogelsberger}}]{pillepichFirstResultsIllustrisTNG2018a}
{Pillepich}, A., {Nelson}, D., {Hernquist}, L., {et~al.} 2018{\natexlab{a}},
  \href{http://dx.doi.org/10.1093/mnras/stx3112}{\color{magenta}\mnras},
  \href{https://ui.adsabs.harvard.edu/abs/2018MNRAS.475..648P}{475, 648}

\bibitem[{{Pillepich} {et~al.}(2018{\natexlab{b}}){Pillepich}, {Springel},
  {Nelson}, {Genel}, {Naiman}, {Pakmor}, {Hernquist}, {Torrey}, {Vogelsberger},
  {Weinberger}, \& {Marinacci}}]{pillepichSimulatingGalaxyFormation2018a}
{Pillepich}, A., {Springel}, V., {Nelson}, D., {et~al.} 2018{\natexlab{b}},
  \href{http://dx.doi.org/10.1093/mnras/stx2656}{\color{magenta}\mnras},
  \href{https://ui.adsabs.harvard.edu/abs/2018MNRAS.473.4077P}{473, 4077}

\bibitem[{{Planck Collaboration} {et~al.}(2016){Planck Collaboration}, {Ade},
  {Aghanim}, {Arnaud}, {Ashdown}, {Aumont}, {Baccigalupi}, {Banday},
  {Barreiro}, {Bartlett}, {Bartolo}, {Battaner}, {Battye}, {Benabed},
  {Beno{\^\i}t}, {Benoit-L{\'e}vy}, {Bernard}, {Bersanelli}, {Bielewicz},
  {Bock}, {Bonaldi}, {Bonavera}, {Bond}, {Borrill}, {Bouchet}, {Boulanger},
  {Bucher}, {Burigana}, {Butler}, {Calabrese}, {Cardoso}, {Catalano},
  {Challinor}, {Chamballu}, {Chary}, {Chiang}, {Chluba}, {Christensen},
  {Church}, {Clements}, {Colombi}, {Colombo}, {Combet}, {Coulais}, {Crill},
  {Curto}, {Cuttaia}, {Danese}, {Davies}, {Davis}, {de Bernardis}, {de Rosa},
  {de Zotti}, {Delabrouille}, {D{\'e}sert}, {Di Valentino}, {Dickinson},
  {Diego}, {Dolag}, {Dole}, {Donzelli}, {Dor{\'e}}, {Douspis}, {Ducout},
  {Dunkley}, {Dupac}, {Efstathiou}, {Elsner}, {En{\ss}lin}, {Eriksen},
  {Farhang}, {Fergusson}, {Finelli}, {Forni}, {Frailis}, {Fraisse},
  {Franceschi}, {Frejsel}, {Galeotta}, {Galli}, {Ganga}, {Gauthier}, {Gerbino},
  {Ghosh}, {Giard}, {Giraud-H{\'e}raud}, {Giusarma}, {Gjerl{\o}w},
  {Gonz{\'a}lez-Nuevo}, {G{\'o}rski}, {Gratton}, {Gregorio}, {Gruppuso},
  {Gudmundsson}, {Hamann}, {Hansen}, {Hanson}, {Harrison}, {Helou},
  {Henrot-Versill{\'e}}, {Hern{\'a}ndez-Monteagudo}, {Herranz}, {Hildebrandt},
  {Hivon}, {Hobson}, {Holmes}, {Hornstrup}, {Hovest}, {Huang}, {Huffenberger},
  {Hurier}, {Jaffe}, {Jaffe}, {Jones}, {Juvela}, {Keih{\"a}nen}, {Keskitalo},
  {Kisner}, {Kneissl}, {Knoche}, {Knox}, {Kunz}, {Kurki-Suonio}, {Lagache},
  {L{\"a}hteenm{\"a}ki}, {Lamarre}, {Lasenby}, {Lattanzi}, {Lawrence}, {Leahy},
  {Leonardi}, {Lesgourgues}, {Levrier}, {Lewis}, {Liguori}, {Lilje},
  {Linden-V{\o}rnle}, {L{\'o}pez-Caniego}, {Lubin}, {Mac{\'\i}as-P{\'e}rez},
  {Maggio}, {Maino}, {Mandolesi}, {Mangilli}, {Marchini}, {Maris}, {Martin},
  {Martinelli}, {Mart{\'\i}nez-Gonz{\'a}lez}, {Masi}, {Matarrese}, {McGehee},
  {Meinhold}, {Melchiorri}, {Melin}, {Mendes}, {Mennella}, {Migliaccio},
  {Millea}, {Mitra}, {Miville-Desch{\^e}nes}, {Moneti}, {Montier}, {Morgante},
  {Mortlock}, {Moss}, {Munshi}, {Murphy}, {Naselsky}, {Nati}, {Natoli},
  {Netterfield}, {N{\o}rgaard-Nielsen}, {Noviello}, {Novikov}, {Novikov},
  {Oxborrow}, {Paci}, {Pagano}, {Pajot}, {Paladini}, {Paoletti}, {Partridge},
  {Pasian}, {Patanchon}, {Pearson}, {Perdereau}, {Perotto}, {Perrotta},
  {Pettorino}, {Piacentini}, {Piat}, {Pierpaoli}, {Pietrobon}, {Plaszczynski},
  {Pointecouteau}, {Polenta}, {Popa}, {Pratt}, \&
  {Pr{\'e}zeau}}]{planckcollaborationPlanck2015Results2016a}
{Planck Collaboration}, {Ade}, P.~A.~R., {Aghanim}, N., {et~al.} 2016,
  \href{http://dx.doi.org/10.1051/0004-6361/201525830}{\color{magenta}\aap},
  \href{https://ui.adsabs.harvard.edu/abs/2016A&A...594A..13P}{594, A13}

\bibitem[{{Sinha} \& {Garrison}(2020)}]{sinhaCorrfuncSuiteBlazing2020a}
{Sinha}, M. \& {Garrison}, L.~H. 2020,
  \href{http://dx.doi.org/10.1093/mnras/stz3157}{\color{magenta}\mnras},
  \href{https://ui.adsabs.harvard.edu/abs/2020MNRAS.491.3022S}{491, 3022}

\bibitem[{{Springel}(2010)}]{springelPurSiMuove2010a}
{Springel}, V. 2010,
  \href{http://dx.doi.org/10.1111/j.1365-2966.2009.15715.x}{\color{magenta}\mnras},
  \href{https://ui.adsabs.harvard.edu/abs/2010MNRAS.401..791S}{401, 791}

\bibitem[{{Springel} {et~al.}(2018){Springel}, {Pakmor}, {Pillepich},
  {Weinberger}, {Nelson}, {Hernquist}, {Vogelsberger}, {Genel}, {Torrey},
  {Marinacci}, \& {Naiman}}]{springelFirstResultsIllustrisTNG2018a}
{Springel}, V., {Pakmor}, R., {Pillepich}, A., {et~al.} 2018,
  \href{http://dx.doi.org/10.1093/mnras/stx3304}{\color{magenta}\mnras},
  \href{https://ui.adsabs.harvard.edu/abs/2018MNRAS.475..676S}{475, 676}

\bibitem[{{Springel} {et~al.}(2005){Springel}, {White}, {Jenkins}, {Frenk},
  {Yoshida}, {Gao}, {Navarro}, {Thacker}, {Croton}, {Helly}, {Peacock}, {Cole},
  {Thomas}, {Couchman}, {Evrard}, {Colberg}, \&
  {Pearce}}]{SpringelSimulationsGalaxiesandQuasars2005}
{Springel}, V., {White}, S. D.~M., {Jenkins}, A., {et~al.} 2005,
  \href{http://dx.doi.org/10.1038/nature03597}{\color{magenta}\nat},
  \href{https://ui.adsabs.harvard.edu/abs/2005Natur.435..629S}{435, 629}

\bibitem[{{Springel} {et~al.}(2001){Springel}, {White}, {Tormen}, \&
  {Kauffmann}}]{springelPopulatingClusterGalaxies2001b}
{Springel}, V., {White}, S. D.~M., {Tormen}, G., \& {Kauffmann}, G. 2001,
  \href{http://dx.doi.org/10.1046/j.1365-8711.2001.04912.x}{\color{magenta}\mnras},
  \href{https://ui.adsabs.harvard.edu/abs/2001MNRAS.328..726S}{328, 726}

\bibitem[{{The Dark Energy Survey Collaboration}(2005)}]{des_red2005}
{The Dark Energy Survey Collaboration}. 2005,
  \href{https://ui.adsabs.harvard.edu/abs/2005astro.ph.10346T}{\href{http://dx.doi.org/10.48550/arXiv.astro-ph/0510346}{\color{magenta}arXiv
  e-prints}, astro}

\bibitem[{{The pandas development
  Team}(2020)}]{thepandasdevelopmentteamPandas2024}
{The pandas development Team}. 2020, {pandas-dev/pandas: Pandas}

\bibitem[{{Tucci} {et~al.}(2021){Tucci}, {Montero-Dorta}, {Abramo},
  {Sato-Polito}, \& {Artale}}]{tucci2021spinbias}
{Tucci}, B., {Montero-Dorta}, A.~D., {Abramo}, L.~R., {Sato-Polito}, G., \&
  {Artale}, M.~C. 2021,
  \href{http://dx.doi.org/10.1093/mnras/staa3319}{\color{magenta}\mnras},
  \href{https://ui.adsabs.harvard.edu/abs/2021MNRAS.500.2777T}{500, 2777}

\bibitem[{{Virtanen} {et~al.}(2020){Virtanen}, {Gommers}, {Oliphant},
  {Haberland}, {Reddy}, {Cournapeau}, {Burovski}, {Peterson}, {Weckesser},
  {Bright}, {van der Walt}, {Brett}, {Wilson}, {Millman}, {Mayorov}, {Nelson},
  {Jones}, {Kern}, {Larson}, {Carey}, {Polat}, {Feng}, {Moore}, {VanderPlas},
  {Laxalde}, {Perktold}, {Cimrman}, {Henriksen}, {Quintero}, {Harris},
  {Archibald}, {Ribeiro}, {Pedregosa}, {van Mulbregt}, \& {SciPy 1. 0
  Contributors}}]{virtanenSciPy10Fundamental2020}
{Virtanen}, P., {Gommers}, R., {Oliphant}, T.~E., {et~al.} 2020,
  \href{http://dx.doi.org/10.1038/s41592-019-0686-2}{\color{magenta}Nat.
  Methods}, \href{https://ui.adsabs.harvard.edu/abs/2020NatMe..17..261V}{17,
  261}

\bibitem[{{Wang} {et~al.}(2022){Wang}, {Mao}, {Zentner}, {Guo}, {Lange}, {van
  den Bosch}, \& {Mezini}}]{Wang2022_gab}
{Wang}, K., {Mao}, Y.-Y., {Zentner}, A.~R., {et~al.} 2022,
  \href{http://dx.doi.org/10.1093/mnras/stac2465}{\color{magenta}\mnras},
  \href{https://ui.adsabs.harvard.edu/abs/2022MNRAS.516.4003W}{516, 4003}

\bibitem[{{Waskom}(2021)}]{waskomSeabornStatisticalData2021}
{Waskom}, M. 2021,
  \href{http://dx.doi.org/10.21105/joss.03021}{\color{magenta}JOSS},
  \href{https://ui.adsabs.harvard.edu/abs/2021JOSS....6.3021W}{6, 3021}

\bibitem[{{Wechsler} {et~al.}(2006){Wechsler}, {Zentner}, {Bullock},
  {Kravtsov}, \& {Allgood}}]{wechslerDependenceHaloClustering2006}
{Wechsler}, R.~H., {Zentner}, A.~R., {Bullock}, J.~S., {Kravtsov}, A.~V., \&
  {Allgood}, B. 2006,
  \href{http://dx.doi.org/10.1086/507120}{\color{magenta}\apj},
  \href{https://ui.adsabs.harvard.edu/abs/2006ApJ...652...71W}{652, 71}

\bibitem[{{Weinberger} {et~al.}(2017){Weinberger}, {Springel}, {Hernquist},
  {Pillepich}, {Marinacci}, {Pakmor}, {Nelson}, {Genel}, {Vogelsberger},
  {Naiman}, \& {Torrey}}]{weinbergerSimulatingGalaxyFormation2017a}
{Weinberger}, R., {Springel}, V., {Hernquist}, L., {et~al.} 2017,
  \href{http://dx.doi.org/10.1093/mnras/stw2944}{\color{magenta}\mnras},
  \href{https://ui.adsabs.harvard.edu/abs/2017MNRAS.465.3291W}{465, 3291}

\bibitem[{{W}es {M}c{K}inney(2010)}]{mckinney-proc-scipy-2010}
{W}es {M}c{K}inney. 2010, {D}ata {S}tructures for {S}tatistical {C}omputing in
  {P}ython

\bibitem[{{White} \& {Rees}(1978)}]{TwoStageGalaxyFormationWhiteRees1978}
{White}, S.~D.~M. \& {Rees}, M.~J. 1978,
  \href{http://dx.doi.org/10.1093/mnras/183.3.341}{\color{magenta}\mnras},
  \href{https://ui.adsabs.harvard.edu/abs/1978MNRAS.183..341W}{183, 341}

\bibitem[{{Xu} {et~al.}(2021){Xu}, {Zehavi}, \&
  {Contreras}}]{xuDissectingModellingGalaxy2021a}
{Xu}, X., {Zehavi}, I., \& {Contreras}, S. 2021,
  \href{http://dx.doi.org/10.1093/mnras/stab100}{\color{magenta}\mnras},
  \href{https://ui.adsabs.harvard.edu/abs/2021MNRAS.502.3242X}{502, 3242}

\bibitem[{{York} {et~al.}(2000){York}, {Adelman}, {Anderson}, {Anderson},
  {Annis}, {Bahcall}, {Bakken}, {Barkhouser}, {Bastian}, {Berman}, {Boroski},
  {Bracker}, {Briegel}, {Briggs}, {Brinkmann}, {Brunner}, {Burles}, {Carey},
  {Carr}, {Castander}, {Chen}, {Colestock}, {Connolly}, {Crocker}, {Csabai},
  {Czarapata}, {Davis}, {Doi}, {Dombeck}, {Eisenstein}, {Ellman}, {Elms},
  {Evans}, {Fan}, {Federwitz}, {Fiscelli}, {Friedman}, {Frieman}, {Fukugita},
  {Gillespie}, {Gunn}, {Gurbani}, {de Haas}, {Haldeman}, {Harris}, {Hayes},
  {Heckman}, {Hennessy}, {Hindsley}, {Holm}, {Holmgren}, {Huang}, {Hull},
  {Husby}, {Ichikawa}, {Ichikawa}, {Ivezi{\'c}}, {Kent}, {Kim}, {Kinney},
  {Klaene}, {Kleinman}, {Kleinman}, {Knapp}, {Korienek}, {Kron}, {Kunszt},
  {Lamb}, {Lee}, {Leger}, {Limmongkol}, {Lindenmeyer}, {Long}, {Loomis},
  {Loveday}, {Lucinio}, {Lupton}, {MacKinnon}, {Mannery}, {Mantsch}, {Margon},
  {McGehee}, {McKay}, {Meiksin}, {Merelli}, {Monet}, {Munn}, {Narayanan},
  {Nash}, {Neilsen}, {Neswold}, {Newberg}, {Nichol}, {Nicinski}, {Nonino},
  {Okada}, {Okamura}, {Ostriker}, {Owen}, {Pauls}, {Peoples}, {Peterson},
  {Petravick}, {Pier}, {Pope}, {Pordes}, {Prosapio}, {Rechenmacher}, {Quinn},
  {Richards}, {Richmond}, {Rivetta}, {Rockosi}, {Ruthmansdorfer}, {Sandford},
  {Schlegel}, {Schneider}, {Sekiguchi}, {Sergey}, {Shimasaku}, {Siegmund},
  {Smee}, {Smith}, {Snedden}, {Stone}, {Stoughton}, {Strauss}, {Stubbs},
  {SubbaRao}, {Szalay}, {Szapudi}, {Szokoly}, {Thakar}, {Tremonti}, {Tucker},
  {Uomoto}, {Vanden Berk}, {Vogeley}, {Waddell}, {Wang}, {Watanabe},
  {Weinberg}, {Yanny}, {Yasuda}, \& {SDSS Collaboration}}]{york2000_sdss}
{York}, D.~G., {Adelman}, J., {Anderson}, Jr., J.~E., {et~al.} 2000,
  \href{http://dx.doi.org/10.1086/301513}{\color{magenta}\aj},
  \href{https://ui.adsabs.harvard.edu/abs/2000AJ....120.1579Y}{120, 1579}

\bibitem[{{Yuan} {et~al.}(2020){Yuan}, {Eisenstein}, \&
  {Leauthaud}}]{yuan2020_lilab}
{Yuan}, S., {Eisenstein}, D.~J., \& {Leauthaud}, A. 2020,
  \href{http://dx.doi.org/10.1093/mnras/staa634}{\color{magenta}\mnras},
  \href{https://ui.adsabs.harvard.edu/abs/2020MNRAS.493.5551Y}{493, 5551}

\bibitem[{{Yuan} {et~al.}(2022{\natexlab{a}}){Yuan}, {Garrison}, {Eisenstein},
  \& {Wechsler}}]{Yuan2022_nogab}
{Yuan}, S., {Garrison}, L.~H., {Eisenstein}, D.~J., \& {Wechsler}, R.~H.
  2022{\natexlab{a}},
  \href{http://dx.doi.org/10.1093/mnras/stac1830}{\color{magenta}\mnras},
  \href{https://ui.adsabs.harvard.edu/abs/2022MNRAS.515..871Y}{515, 871}

\bibitem[{{Yuan} {et~al.}(2022{\natexlab{b}}){Yuan}, {Garrison}, {Hadzhiyska},
  {Bose}, \& {Eisenstein}}]{Yuan2022_yesgab}
{Yuan}, S., {Garrison}, L.~H., {Hadzhiyska}, B., {Bose}, S., \& {Eisenstein},
  D.~J. 2022{\natexlab{b}},
  \href{http://dx.doi.org/10.1093/mnras/stab3355}{\color{magenta}\mnras},
  \href{https://ui.adsabs.harvard.edu/abs/2022MNRAS.510.3301Y}{510, 3301}

\bibitem[{{Yuan} {et~al.}(2021){Yuan}, {Hadzhiyska}, {Bose}, {Eisenstein}, \&
  {Guo}}]{Yuan2021_gab}
{Yuan}, S., {Hadzhiyska}, B., {Bose}, S., {Eisenstein}, D.~J., \& {Guo}, H.
  2021, \href{http://dx.doi.org/10.1093/mnras/stab235}{\color{magenta}\mnras},
  \href{https://ui.adsabs.harvard.edu/abs/2021MNRAS.502.3582Y}{502, 3582}

\bibitem[{{Zehavi} {et~al.}(2018){Zehavi}, {Contreras}, {Padilla}, {Smith},
  {Baugh}, \& {Norberg}}]{zehaviImpactAssemblyBias2018a}
{Zehavi}, I., {Contreras}, S., {Padilla}, N., {et~al.} 2018,
  \href{http://dx.doi.org/10.3847/1538-4357/aaa54a}{\color{magenta}\apj},
  \href{https://ui.adsabs.harvard.edu/abs/2018ApJ...853...84Z}{853, 84}

\bibitem[{{Zentner}(2007)}]{zentnerExcuersionSetTheory2007}
{Zentner}, A.~R. 2007,
  \href{http://dx.doi.org/10.1142/S0218271807010511}{\color{magenta}International
  Journal of Modern Physics D},
  \href{https://ui.adsabs.harvard.edu/abs/2007IJMPD..16..763Z}{16, 763}

\bibitem[{{Zentner} {et~al.}(2014){Zentner}, {Hearin}, \& {van den
  Bosch}}]{zentnerGalaxyAssemblyBias2014a}
{Zentner}, A.~R., {Hearin}, A.~P., \& {van den Bosch}, F.~C. 2014,
  \href{http://dx.doi.org/10.1093/mnras/stu1383}{\color{magenta}\mnras},
  \href{https://ui.adsabs.harvard.edu/abs/2014MNRAS.443.3044Z}{443, 3044}

\bibitem[{{Zhao} {et~al.}(2025){Zhao}, {Peng}, {Jing}, {Yang}, {Ho}, {Renzini},
  {Gallazzi}, {Lyu}, {Maiolino}, {Dou}, {Gao}, {Gu}, {Mannucci}, {Mo}, {Wang},
  {Wang}, {Wang}, {Wang}, {Xu}, {Yuan}, \& {Zhu}}]{zhao2025_halomass}
{Zhao}, D., {Peng}, Y., {Jing}, Y., {et~al.} 2025,
  \href{http://dx.doi.org/10.3847/1538-4357/ad991f}{\color{magenta}\apj},
  \href{https://ui.adsabs.harvard.edu/abs/2025ApJ...979...42Z}{979, 42}

\bibitem[{{Zu} \& {Mandelbaum}(2018)}]{zu2018MappingStellarContent}
{Zu}, Y. \& {Mandelbaum}, R. 2018,
  \href{http://dx.doi.org/10.1093/mnras/sty279}{\color{magenta}\mnras},
  \href{https://ui.adsabs.harvard.edu/abs/2018MNRAS.476.1637Z}{476, 1637}

\end{thebibliography}

\end{document}